\documentclass[review]{elsarticle}
 \biboptions{comma,sort&compress}

\usepackage[table]{xcolor}

\usepackage{graphicx}
\usepackage{amsmath}
\usepackage{here}
\usepackage{cuted}
\usepackage{here}
\usepackage[dvips]{epsfig}
\definecolor{bluette}{rgb}{.2,.4,0}
\definecolor{salmon}{rgb}{.9,0.68,0.5}
\definecolor{motive}{rgb}{0.2,1,.5}
\definecolor{list}{rgb}{0.3,.8,.1}
\definecolor{moe}{rgb}{1,.7,.5}
\definecolor{mote}{rgb}{.7,.5,.6}
\definecolor{pisello}{rgb}{.1,1,0}
\definecolor{orange}{rgb}{1,.7,0}
\definecolor{oliva}{rgb}{.1,.5,0.3}
\definecolor{greenda}{rgb}{0,.3,.2}
\definecolor{greenli}{rgb}{0.5,.8,.0}
\definecolor{blueda}{rgb}{0,.1,.6}
\definecolor{purple}{rgb}{.7,.1,.2}
\definecolor{marrone}{rgb}{1,0.7,0}
\definecolor{pinky}{rgb}{1,0.8,0.8}
\definecolor{rose}{rgb}{1,0.4,0}

\def\oliva{\color{oliva}}

\def\beq{\begin{equation}}
\def\eeq{\end{equation}}
\def\bea{\begin{eqnarray}}
\def\eea{\end{eqnarray}}
\def\bq{\begin{quote}}
\def\eq{\end{quote}}

\def\nnb{\nonumber}
\def\ga{\left(}
\def\dr{\right)}

\def\rar{\rightarrow}
\def\lrar{\Longrightarrow}
\def\lrar2{\longrightarrow}
																
\def\nnb{\nonumber}

\def\la{\langle}
\def\ra{\rangle}

\def\ba{\vspace*{-0.2cm}\begin{array}}
\def\ea{\end{array}\vspace*{-0.2cm}}

\def\b{$\bullet~$}
\def\d{$\diamond~$}

\def\als{\alpha_s}

\def\gg2{\la\alpha_s G^2 \ra}
\def\gg3{g^3f_{abc}\la G^aG^bG^c \ra}
\def\ggg4{\la\als^2G^4\ra}

\def\gg{\lag g^{2}_{s} G^2 \rag}
\def\ggg{\lag g^{3}_{s}G^3\rag}
\usepackage{amsmath}
\usepackage{slashed}
\usepackage{color}

\begin{document}
%\markboth{Stephan Narison, Montpellier (FR)}{ }
\begin{frontmatter}

%\title{QCD condensates, $\alpha_s$ and $a_\mu$  from $e^+e^-\to$ I=1 Hadrons data}
\title{QCD parameters and SM-high precision from $e^+e^-\to$  Hadrons\,: Updated\,\tnoteref{invit}}
\tnotetext[invit]{This is an update of the results in Ref.\,\cite{SNe}, where some preliminary ones have been presented  at the alphas 2024 workshop (5-9 february 2024, ECT* Trento-IT).}
\author{Stephan Narison%\corref{cor1}
}
\address{Laboratoire
Univers et Particules de Montpellier (LUPM), CNRS-IN2P3, \\
Case 070, Place Eug\`ene
Bataillon, 34095 - Montpellier, France\\
and\\
Institute of High-Energy Physics of Madagascar (iHEPMAD)\\
University of Ankatso, Antananarivo 101, Madagascar}
\ead{snarison@yahoo.fr}

%%
%%%%%%%%%%%%%%%%%%%%%%%%%%%%%%%%%%%%%%%%%%%%%%%%%

\date{\today}
\begin{abstract}
%\noindent

{\bf 1.} I report an update of my previous  comparison of the theoretical  value of the muon anomaly $a_\mu\equiv \frac{1}{2}(g-2)_\mu$ with the new measurement. One finds\,: $\Delta a_\mu\equiv a_\mu^{exp}-a_\mu^{th} = (143\pm 42_{th}\pm 22_{exp})\times 10^{-11}$ indicating about 3$\sigma$ discrepancy  between the SM predictions and experiment. 

{\bf 2.}  I improve the estimate  of  QCD power corrections up to dimension $D=12$  and provide a new estimate of the ones up to $D=20$ within  the  Shifman-Vainshtein-Zahkarov (SVZ) expansion by combining the ratio of the SVZ Borel/Laplace sum rules (LSR) with the Braaten-Pich and the author (BNP) $\tau$-like decay moments\,
 for the $I=1$ vector current. The results  summarized in Table\,1 confirm a violation of the factorization of  the four-quark condensates and the value of the gluon one $\la \alpha_s G^2\ra$ from some other sources. Up to $D=20$, I do not observe any factorial nor exponential growth of the size of these power corrections. 

{\bf 3.} I use these new values of the $D=6,8$ power corrections to extract  $\alpha_s$ from the BNP lowest moment. To order $\alpha_s^4$, I find within the SVZ expansion\,: 
$\alpha_s(M_\tau)\vert^{SVZ}_{e^+e^-}= 0.3081(49)_{fit}(71)_{\alpha_s^5}$ [resp.  $0.3260 (47)_{fit}(62)_{\alpha_s^5}]$  implying
$\alpha_s(M_Z)\vert^{SVZ}_{e^+e^-}= 0.1170(6)(3)_{evol}$
[resp. $0.1192(6)(3)_{evol}$] for Fixed Order (FO) [resp. Contour Improved (CI)] PT series. 
They  lead to  the mean: 
$ \alpha_s(M_\tau)\vert^{SVZ}_{e^+e^-}=0.3179(58)_{fit}(81)_{syst}$ and 
$ \alpha_s(M_Z)\vert^{SVZ}_{e^+e^-}= 0.1182(12)(3)_{evol}$ where the systematic error(syst)  takes into account the discrepancy between the FO and CI results. 
Using  the lowest BNP moment, we obtain from the vector (V) component of $\tau$-decay data: $ \alpha_s(M_\tau)\vert_{\tau,V}^{SVZ}=0.3128(19)_{fit}(77)_{\alpha_s^5}$ [resp.  $0.3291 (25)_{fit}(65)_{\alpha_s^5}]$   implying
$\alpha_s(M_Z)\vert_{\tau,V}^{SVZ}= 0.1176(10)(3)_{evol}$
[resp. $0.1196(8)(3)_{evol}$] for FO [resp. CI] PT series,
giving the mean:  $\alpha_s(M_\tau)\vert_{\tau,V}^{SVZ}=0.3219(52)(91_{syst}$ leading to: $\alpha_s(M_Z)\vert_{\tau,V}^{SVZ}=0.1187(13)(3)_{evol}$. The average of the two determinations from $e^+e^-$ and $\tau$-decay data is:  $ \la\alpha_s(M_\tau)\ra=0.3198(72)$ which implies  $ \la\alpha_s(M_Z)\ra= 0.1185(9)(3)_{evol}$. 

{\bf 4.} Some (eventual) contributions beyond the SVZ expansion ($1/Q^2$, instantons and duality violation) are discussed in Sections 10 and 11 which are expected to be relatively small. 

%\ccode{Pac numbers: 11.55.Hx, 12.38.Lg, 13.20-Gd, 14.65.Dw, 14.65.Fy, 14.70.Dj}  
\end{abstract}
\end{frontmatter}
%\end{document}
%%%%%%%%%%%%%%%%%%%%%%%%%%%%%%%%%%
\newpage
%%%%%%%%%%%%%%%%%%%%%%%%%%%%%%%%%%
%\vspace*{-1.5cm}
\section{Introduction}
\vspace*{-0.2cm}
 %\nin
%%%%%%%%%%%%%%%%%%%%%%%%%%%%%%%%%%%
Precise determinations of the hadronic contributions to the muon anomaly $a_\mu\equiv \frac{1}{2}(g-2)_\mu$ and the QCD parameters (power corrections and $\alpha_s$)  are important inputs  for testing the Standard Model (SM) and for QCD and hadrons phenomenology. 

In this paper, I shall revisit the results obtained in\,\cite{SNe} (referred hereafter as SN23) due to the new experimental measurement of the positive charged muon anomaly\,\cite{MG2} received after the publication of the papers in SN23. 

 I shall also re-estimate the QCD condensates and some higher dimension ones by combining  the analysis from  the ratio of the SVZ\,\cite{SVZ}\,\footnote{For reviews, see e.g.\,\cite{ZAKA, SNB1,SNB2,SNREV1} and references quoted in\,\cite{SNe}.} \,LSR\,\cite{SNR,BELL}\footnote{For a recent review, see e.g.\,\cite{SNLSR}.} with the one from some higher BNP\,$\tau$-like moments\,\cite{BNP2,BNP,LEDI}. These new values of QCD condensates will be used as inputs for determining $\alpha_s$ from the BNP lowest moment.

In so doing, I shall use, like in SN23\,\cite{SNe}, the PDG 22 compilation of the $e^+e^-\to$  Hadrons\,\cite{PDG}  $\oplus$ the recent CMD3 data\,\cite{CMD3} for the pion form factor and the value of gluon condensate $\la\alpha_s G^2\ra$ from heavy quarkonia and some other sources.

% we have obtained in Ref.\,\cite{SNe} (hereafter referred as SN23) some predictions for the dimension $D=4,6,8$ condensates using the ratio of of Laplace Sum Rules (LSR) . However, these results may not be robust as they are obtained at relatively large value of the sum rule scale $\tau\simeq2$ GeV$^{-2}$ where the convergence of the OPE is questionable. 

\vspace*{-0.45cm}

%%%%%%%%%%%%%%%%%%%%%%%%%%%%%%%%
\section{The $I=1$  isovector two-point function}
%%%%%%%%%%%%%%%%%%%%%%%%%%%%%%%%%
We shall be concerned with the two-point correlator :
%\vspace*{-0.5cm}
\beq
\hspace*{-0.6cm} 
\Pi^{\mu\nu}_H(q^2)=i\hspace*{-0.1cm}\int \hspace*{-0.15cm}d^4x ~e^{-iqx}\la 0\vert {\cal T} {J^\mu_H}(x)\ga {J^\nu_H}(0)\dr^\dagger \vert 0\ra %\nnb\\
~~=-(g^{\mu\nu}q^2-q^\mu q^\nu)\Pi_H(q^2)
%\int_{t_>}^\infty \frac{dt}{t-q^2-i\epsilon}\frac{1}{\pi}{\rm Im}\Pi_H(t)+\cdots,
 \label{eq:2-point}
 \eeq
built from the T-product of the bilinear $I=1$  vector current :
\beq
 J^\mu_H(x)=\frac{1}{2}{[}: \bar\psi_u\gamma^\mu\psi_u-\bar\psi_d\gamma^\mu\psi_d:{]}.
\eeq
It obeys the dispersion relation:
\beq
\Pi_H(q^2)=\int_{t>}^\infty \frac{dt}{t-q^2-i\epsilon} \frac{1}{\pi}\,{\rm Im} \Pi_H(t)+\cdots,
\eeq
where $\cdots$ are subtraction constants polynomial in $q^2$ and $t>$   is the hadronic threshold..
%%%%%%%%%%%%%%%%%%%%%%%%%%%%%%%%
%\subsection*{\b The spectral function from the isovector $e^+e^-\to\,{\rm Hadrons}$ data}
%%%%%%%%%%%%%%%%%%%%%%%%%%%%%%%%%
From the optical theorem, the spectral function ${\rm Im} \Pi_H(t)$ can be related to the $e^+e^-\to\,{\rm Hadrons}$
total cross-section as:
\beq
R^{I=1}_{ee}\equiv \frac{\sigma(e^+e^-\to\,{\rm Hadrons}}{\sigma(e^+e^-\to\mu^+\mu^-)}=\ga\frac{3}{2}\dr 
8\pi\, {\rm Im} \Pi_H(t).
\eeq
The data handlings have been discussed in details  in Ref.\,\cite{SNe} and will not be repeated here. We use the compilation of PDG22\,\cite{PDG} which take into account the correlations among different data while we use the new CMD3\,\cite{CMD3} data for the pion form factor below 1 GeV. In this paper, we shall be concerned with the data below 1.875 GeV.
%%%%%%%%%%%%%%%%%%%%%%%%%%%%%%%%%%%%%%%%%%%%%%%%%
\section{Comparison of the experimental and theoretical values of the muon anomaly $a_\mu$}
%%%%%%%%%%%%%%%%%%%%%%%%%%%%%%%%%%%%%%%%%%%%%%%%%
We start the paper by updating our previous comparison (1st reference in Ref.\,\cite{SNe})  of the theoretical and new experimental values of the muon anomaly $a_\mu$\,\cite{MG2}\,:
\beq
a_\mu^{exp}= 116592059(22)\times 10^{-11},
\eeq
which improves the accuracy of previous  results in Refs.\,\cite{BNL} and FNAL\,\cite{FNAL} \, by a factor 1.86. 

This experimental value is to be compared with the theoeretical results compiled in Table 1 of Ref.\,\cite{GM2} and in  Table 8 of Ref.\,\cite{KNECHT}  to which we add the new estimate of the lowest order hadronic contributions to the vacuum polarization:
\beq
a_\mu\vert^{hvp}_{l.o}=(7036.5\pm 38.9)\times 10^{-11}.
\eeq
We notice that our prediction using the CMD3 data is larger by an amount of about $100\times 10^{-11}$ compared to the previous estimates using KLOE data as compared in Fig. 17 of SN23\,\cite{SNe} and the discussion in Section 10 of this paper. 
Therefore, we obtain\,(1st reference in Ref.\,\cite{SNe})\,:
\beq
a_\mu^{th}= 116591916(42)\times 10^{-11}.
\eeq
This leads to (2nd reference in Ref.\,\cite{SNe})\,:
\beq
\Delta a_\mu\equiv a_\mu^{exp}-a_\mu^{th} = (143\pm 42_{th}\pm 22_{exp})\times 10^{-11},%= (143\pm 47)\times 10^{-11},
\label{eq:amu-sm}
\eeq
which indicates about $3\,\sigma$ discrepancy between experiment and the SM predictions. 
%%%%%%%%%%%%%%%%%%%%%%%%%%%%%%%%
\section{The standard SVZ expansion}
%%%%%%%%%%%%%%%%%%%%%%%%%%%%%%%%%
\vspace*{-0.2cm}
Within the SVZ\,\cite{SVZ} Operator Product Expansion (OPE),  QCD condensates with higher and higher dimensions are assumed to approximate the not yet known QCD non-perturbative contributions.  The SVZ-expansion reads ($q^2\equiv -Q^2$)\,:
\beq
8\pi^2\Pi_H(-Q^2,m_q^2,\mu)=\sum_{D=0,2,..}\hspace*{-0.25cm}\frac{C_{D}(-Q^2,m_q^2,\mu)\la O_{D}(\mu)\ra}{(Q^2)^{D/2}}~, 
\label{eq:ope}
\eeq
where $m_q$ is the quark mass, $\mu$ is the subtraction scale which separates the long and short distance dynamics. $C_{D}$ are perturbatively calculable Wilson coefficients while $\la O_{D}(\mu)\ra$ are non-perturbative QCD condensates of dimension $D$. In the phenomenological analysis, the OPE is often truncated at $D=6,8$ where the approach gives a satisfactory explanation of different data (see e.g. \cite{ZAKA, SNB1,SNB2,SNREV1,SNR,BELL}). 

In addition, one should note that the contributions of higher dimension condensates are not under a good control due to the large number of Feynman diagram ones, to the inaccurate estimate of their size and to the difficulty to built a renormalization group invariant (RGI) condensate due to their mixing under renormalization\,\cite{SNTARRACH}. 

In SN23, one uses the PT expression up to order $\alpha_s^4$ while the OPE is truncated at $D=6,8$. 

One should mention that, besides the well-known $\la\bar \psi\psi\ra$ quark condensate, the gluon condensates have been determined from the heavy quark mass-splittings and some other sum rules\,\cite{SNparam, SNcb1}:
\beq
\la\alpha_s G^2\ra =  (6.39\pm 0.35)\times 10^{-2}\,{\rm GeV^4},~~~~~~~~~{\la g^3  G^3\ra}/{\la\alpha_s G^2\ra}=8.2(1.0)\,[{\rm GeV^4}],
\label{eq:asg2}
\eeq
while different analysis of the light meson systems lead to the value of the four-quark condensate\,(see e.g. the papers quoted in \cite{SNREV1} and \cite{SNe}) :
\beq
\rho \alpha_s\la \bar \psi\psi\ra^2=5.8(9)\times 10^{-4} \,[\rm GeV^6].
\eeq
%%%%%%%%%%%%%%%%%%%%%%%%%%%%%%%%
\section{The   Laplace sum rules (LSR) and their ratios}
%%%%%%%%%%%%%%%%%%%%%%%%%%%%%%%%%
\subsection*{\b Form of the  LSR and their ratio}
%%%%%%%%%%%%%%%%%%%%%%%%%%%
In Ref.\,\cite{LNT} and SN23\,\cite{SNe}, the dimension  $D=4,6$ and 8 condensates appearing in the OPE of the two-point vector correlator have been estimated using the ratio of Laplace sum rule moments\,\cite{SVZ,SNR,BELL}\footnote{For a recent review, see e.g.\,\cite{SNLSR}.}:
\beq
 {\cal R}^c_{10}(\tau)\equiv\frac{{\cal L}^c_{1}} {{\cal L}^c_0}= \frac{\int_{t>}^{t_c}dt~e^{-t\tau}t\, R_{ee}^{I=1}(t,\mu) }   {\int_{t>}^{t_c}dt~e^{-t\tau} R_{ee}^{I=1}(t,\mu) },
\label{eq:lsr}
\eeq
where $\tau$ is the LSR variable, $t>$   is the hadronic threshold.  Here $t_c$ is  the threshold of the ``QCD continuum" which parametrizes, from the discontinuity of the Feynman diagrams, the spectral function  ${\rm Im}\,\Pi_H(t,m_q^2,\mu^2)$.  $m_q$ is the quark mass and $\mu$ is an arbitrary subtraction point. The exponential weight enhances the contribution of the low-energy region of the spectral function accessible experimentally. We shall see in Eq.\ref{eq:svz} that it introduces a factorial factor in front of each condensates which accelerates the convergence of the OPE. 
% The spectral function is related  through the optical theorem to the isovector part of the ratio $R^{I=1}_{ee}$ as:

%%%%%%%%%%%%%%%%%%%%%%%%%%%%%%%%%
\subsection*{\b The PT QCD expression of the  LSR}
%%%%%%%%%%%%%%%%%%%%%%%%%%%%%%%%%%
To order $\alpha_s^4$, the perturbative (PT) expression of the lowest moment reads\,\cite{KAHN}:
\beq
{\cal L}^{PT}_0(\tau)= \frac{3}{2}\tau^{-1}\Big{[} 1+a_s+2.93856\,a_s^2+ 6.2985\,a_s^3 + 22.2233\,a_s^4\Big{]}.
\eeq
Then, taking its derivative in $\tau$, one gets ${\cal L}_1(\tau)$ and then their ratio ${\cal R}_{10}(\tau)$. 
%%%%%%%%%%%%%%%%%%%%%%%%%%%%%%%%%
\subsection*{\b The NPT QCD expression of the  LSR}
%%%%%%%%%%%%%%%%%%%%%%%%%%%%%%%%%%
From Eq.\,\ref{eq:ope}, one can deduce the lowest moment LSR:
\beq
{\cal L}^{NPT}_0(\tau) = \frac{3}{2}\tau^{-1}\sum_D \frac{d_D}{(D/2-1)!} \tau^{D/2} ~,
\label{eq:svz}
\eeq
from which one can deduce ${\cal L}^{NPT}_1$ and ${\cal R}_{10}$ where $d_D\equiv C_D\la O_D\ra$. 
%%%%%%%%%%%%%%%%%%%%%%%%%%%%%%%%
\section{The $\tau$-decay-like moment sum rules}
%%%%%%%%%%%%%%%%%%%%%%%%%%%%
\subsection*{\b The BNP lowest moment}
%%%%%%%%%%%%%%%%%%%%%%%%%%%
  %%%%%%%%%%%%%%%%%%%%%%%%%%%%%%%%%%%%%%%
  %\vspace*{-0.5cm}
\begin{figure}[hbt]
\begin{center}
%{\bf a)}\hspace*{8cm} {\bf b)}\\
\includegraphics[width=8cm]{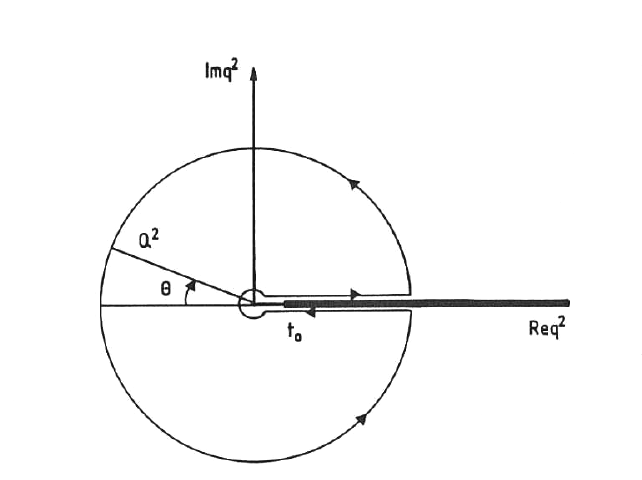}
\caption{\footnotesize Cauchy contour integral in the complex $Q^2\equiv |q^2|$-plane}\label{fig:cauchy}
\end{center}
%\vspace*{-0.5cm}
\end{figure}  
%%%%%%%%%%%%%%%%%%%%%%%%%%%%%%%%%%%%%%%%%%
Initially introduced by BNP\,\cite{BNP,BNP2}, the lowest original $\tau$-decay moment reads in the massless quark limit\,:
\beq
{\cal R}^{ee}_{0}= 12\pi\int_{t>}^{x_0}\hspace*{-0.25cm} dt\,(1-x_0)^2(1+2x_0)\, 2\,R_{ee}^{I=1}(x_0),
\eeq
where : $x_0\equiv t/M_0^2$ with $M_0$ is the $\tau$-like lepton mass. 
The moment can be conveniently reformulated in terms of the Cauchy integral along the contour in Fig.\,\ref{fig:cauchy} \`a la Shankar\,\cite{SHANKAR}\,\footnote{Some other applications in QCD can e.g. by found in\,\cite{FNR}.}:
\beq
{\cal R}^{ee}_{0}= 6i\,\pi\int_{|x_0|=1} \hspace*{-0.5cm} dx_0\, (1-x_0)^2(1+2x_0)\, 2\,R_{ee}^{I=1}(x_0),
\eeq
with a radius $|x_0|=1$. 
To order $\alpha_s^4$, the PT corrections to the lowest BNP moment reads\,\cite{BNP,LEDI,LARIN,CHET4}:
 \bea
\delta^{(0)}_{0}  &=& a_s+5.2026\,a_s^2+26.368\,a_s^3+127.085\,a_s^4+{\cal O}(a_s^5)~~~~~{\rm~~~~(FO)} \nnb\\
&=&1.364\,a_s+2.54\,a_s^2+9.71\,a_s^3+64.29\,a_s^4+{\cal O}(a_s^5)~~~~~{\rm ~~~~(CI)}
\eea
for Fixed Order (FO) and Contour Improved (CI) PT series. 

Unlike different Finite Energy Sum Rule (FESR) used in the current literature\,\cite{FESR} and its pinched versions\,\cite{BOITO}, the main advantage of the $\tau$-decay-like moment is the presence of the weight factor $(1-x_0)^2$ which suppresses the contribution near the real axis where the data are inaccurate. This approach has lead to an accurate determination of $\alpha_s$ at the $\tau$ mass. 

%%%%%%%%%%%%%%%%%%%%%%%%%%%%%%%%
%%%%%%%%%%%%%%%%%%%%%%%%%%%%
\subsection*{\b The $\tau$-like  high moments}
%%%%%%%%%%%%%%%%%%%%%%%%%%%
Inspired by the original work of\,\cite{LEDI}, different forms of high-moments have been proposed in the
literature for extracting simultaneously $\alpha_s$ and the QCD condensates\,\cite{ALEPH,OPAL,PICH1,DAVIER}. In this paper, we shall be concerned with the moments of the form:
\beq
{\cal R}_n^{ee}= 6i\,\pi\int_{|x_0|=1}\hspace*{-0.5cm} dx_0\, (1-x_0)^2(1+2x_0)(x_0^n)\, 2\,R_{ee}^{I=1}(x_0),
\eeq
where $n=1$ to 6.
%%%%%%%%%%%%%%%%%%%%%%%%%%%%%%%%%%
\subsection*{\b Perturbative corrections to the $\tau$-like  high moments at order $\alpha_s^4$}
%%%%%%%%%%%%%%%%%%%%%%%%%%%%%%%%%5
The perturbative contributions read to order $\alpha_s^4$ for fixed order (FO) perturbation series:
\bea
\delta^{(0)}_{1}  &=& a_s+3.7776\,a_s^2+6.8810\,a_s^3-66.7235\,a_s^4+{\cal O}(a_s^5)\nnb\\
\delta^{(0)}_{2} &=& a_s+3.2151\,a_s^2+1.0687\,a_s^3-98.2293\,a_s^4+{\cal O}(a_s^5)\nnb\\
\delta^{(0)}_{3}&=& a_s+2.8990\,a_s^2-1.7710\,a_s^3-108.034\,a_s^4+{\cal O}(a_s^5)\nnb\\
\delta^{(0)}_{4} &=& a_s+2.6928\,a_s^2-3.4664\,a_s^3-111.868\,a_s^4+{\cal O}(a_s^5)\nnb\\
\delta^{(0)}_{5} &=& a_s + 2.4365\,a_s^2 - 5.4048\,a_s^3 - 119.087\,a_s^4+{\cal O}(a_s^5)\nnb\\
\delta^{(0)}_{6} &=& a_s +2.5464\,a_s^2 - 4.5966\,a_s^3 - 108.552\,a_s^4+{\cal O}(a_s^5)
\eea

%%%%%%%%%%%%%%%%%%%%%%%%%%%%%%%%%%
\subsection*{\b $\tau$-like  high moments with non-Perturbative corrections  at lowest order for $m_{u,d}=0$}
%%%%%%%%%%%%%%%%%%%%%%%%%%%%%%%%%5
To lowest order in $\alpha_s$ and in the massless quark limit, the non-perturbative 
corrections to the different moments give:
\bea
 {\cal R}^{ee}_{0}&=&\frac{3}{2}\Bigg{[} 1+\delta^{(0)}_{0} -6\frac{d_6}{M_0^6}-{4}\frac{d_8}{M_0^8}+\sum_D{\delta_0^{(D)}}\Bigg{]}: ~~~~~~~~~~~~~~~~~~~~~~~  ~~~~~~~~\delta^{(D)}_{0}=0 ~~~ (for~~ D\geq 10),\nnb\\
 {\cal R}^{ee}_{1}&=&\frac{9}{20}\Bigg{[} 1+\delta^{(0)}_{1} -\frac{20}{3}\frac{d_4}{M_0^4}+20\frac{d_8}{M_0^8}+\frac{40}{3}\frac{d_{10}}{M_0^{10}}+\sum_D{\delta_1^{(D)}}\Bigg{]}: ~~~~  ~~~~~~~~\delta^{(D)}_{1}=0 ~~~ (for~~ D\geq 12),\nnb\\
 {\cal R}^{ee}_{2}&=&\frac{1}{5}\Bigg{[} 1+\delta^{(0)}_{2} +15\frac{d_6}{M_0^6}-45\frac{d_{10}}{M_0^{10}}-30\frac{d_{12}}{M_0^{12}}+\sum_D{\delta_2^{(D)}}\Bigg{]}: ~~~~ ~~~~~~~~~~\delta^{(D)}_{2}=0 ~~~ (for~~ D\geq 14),\nnb\\
 {\cal R}^{ee}_{3}&=&\frac{3}{28}\Bigg{[} 1+\delta^{(0)}_{3} -28\frac{d_8}{M_0^8}+84\frac{d_{12}}{M_0^{12}}+56\frac{d_{14}}{M_0^{14} }+\sum_D{\delta_3^{(D)}}\Bigg{]}: ~~~~~  ~~~~~~~~\delta^{(D)}_{3}=0 ~~~ (for~~ D\geq 16),\nnb\\
 {\cal R}^{ee}_{4}&=&\frac{9}{140}\Bigg{[} 1+\delta^{(0)}_{4}+\frac{140}{3}\frac{d_{10}}{M_0^{10}}-140\frac{d_{14}}{M_0^{14}}-\frac{280}{3}\frac{d_{16}}{M_0^{16}} +\sum_D{\delta_4^{(D)}} \Bigg{]}: ~~~~ \delta^{(D)}_{4}=0 ~~~ (for~~ D\geq 18), \nnb\\
  {\cal R}^{ee}_{5}&=&\frac{1}{24}\Bigg{[} 1+\delta^{(0)}_{5}-72\frac{d_{12}}{M_0^{12}}-216\frac{d_{16}}{M_0^{16}}-144\frac{d_{18}}{M_0^{18}}  +\sum_D{\delta_5^{(D)}}\Bigg{]}: ~~~~ ~~~~\delta^{(D)}_{5}=0 ~~~ (for~~ D\geq 20), \nnb\\
  {\cal R}^{ee}_{6}&=&\frac{1}{35}\Bigg{[} 1+\delta^{(0)}_{6}+75\frac{d_{14}}{M_0^{14}}-225\frac{d_{18}}{M_0^{18}}-150\frac{d_{20}}{M_0^{20}} +\sum_D{\delta_6^{(D)}} \Bigg{]}: ~~~~ ~~~~\delta^{(D)}_{6}=0 ~~~ (for~~ D\geq 22), \nnb\\
  \eea
  where $\delta_n^{(D)}$ is the contribution of the condensate of dimension $D$ and $n$ is the degree of the moment. 
  %%%%%%%%%%%%%%%%%%%%%%%%%%%%%%%%
\section{Extraction of the QCD condensates}
%%%%%%%%%%%%%%%%%%%%%%%%%%%%
\subsection*{\b Condensates from the $\tau$-decay-like  high moments}
%%%%%%%%%%%%%%%%%%%%%%%%%%%%%%%%
In so doing we fix the value of $\alpha_s$ from the world average\,\cite{PDG}. Then, we are looking for stability in the change of $M_0$ for extracting the optimal values of the condensates. 
%%%%%%%%%%%%%%%%%%%%%%%%%
\subsubsection*{\hspace*{0.5cm} \d $ {\cal R}^{ee}_{0}$ moment}
%%%%%%%%%%%%%%%%%%%%%%%%%
-- Using a two-parameter ($d_6,d_8$) fit in the lowest BNP moment, we do not find a conclusive result.  We interpret this fact due to the opposite sign of the two contributions which tend to cancel out. However,  this feature is welcome for determining $\alpha_s$ as the sum of the non-perturbative contributions are small but may indicate that it is not a good place for determining accurately the condensates. 

-- We impose more constraints by using a one-parameter fit  with as an initial value the one \,:
\beq
d_6= -(20.5\pm 2)\times 10^{-2} ~{\rm GeV}^6,
\label{eq:d6}
\eeq
%$
%\la \alpha_s G^2\ra 
%$ in Eq.\,\ref{eq:asg2} 
%= (6.49\pm 0.35)\times 10^{-2}~{\rm GeV}^4,
%\label{eq:
%\eeq
obtained recently in SN23\,\cite{SNe} from the ratio of LSR into $ {\cal R}^{ee}_{0}$ in an attempt to fix $d_8$.  The resulting value of $d_8$ does not show any $M_0$ stability as also observed in SN23.  Then, we conclude that the analysis from ${\cal R}_0^{ee}$ is not also conclusive. 

%%%%%%%%%%%%%%%%%%%%%%%%%%%%%%%%%%
\subsubsection*{\hspace*{0.5cm} \d   $ {\cal R}^{ee}_{1}$ moment}
%%%%%%%%%%%%%%%%%%%%%%%%%%%%%%%%%%
In the following we shall only work with a one-parameter fit $d_0$ and  use as initial values of $\la \alpha_s G^2\ra $ in Eq.\,\ref{eq:asg2}  obtained from heavy quarks mass-splittings and some other sources  and the value of $d_8$ from SN23:
\beq
d_8 = (4.7\pm 3.5)\times 10^{-2}~{\rm GeV}^8.
\label{eq:d8}
\eeq
We extract $d_{10}$. Then, we repeat the analysis by fixing $d_{10}$ and $d_{6}$ and  deduce $d_8$.  
%%%%%%%%%%%%%%%%%%%%%%%%%
\subsubsection*{\hspace*{0.5cm} \d   $ {\cal R}^{ee}_{2}$ moment}
%%%%%%%%%%%%%%%%%%%%%%%%%
 We use the value of $d_{10}$ into $ {\cal R}^{ee}_{2}$ and extract $d_{12}$. Using this value of $d_{12}$, we re-iterate the analysis to extract $d_{10}$ and continue the iterations. 

%%%%%%%%%%%%%%%%%%%%%%%%%
\subsubsection*{\hspace*{0.5cm} \d  Higher moments  $n=2$ to 6}
%%%%%%%%%%%%%%%%%%%%%%%%%
 We repeat the procedure for higher moments and extract $d_{14}$ to $d_{20}$. The quoted values are obtained without any iterations as they are already quite precise. However, the quoted error does not include the systematic ones which may increase with the degree of moments.

%%%%%%%%%%%%%%%%%%%%%%%%%
\subsubsection*{\hspace*{0.5cm} \d  Conclusions from $\tau$-like moments}
\begin{figure}[hbt]
\begin{center}
\hspace*{-4cm}{\bf a)}\hspace*{8cm} {\bf b)}\\
\includegraphics[width=7.5cm]{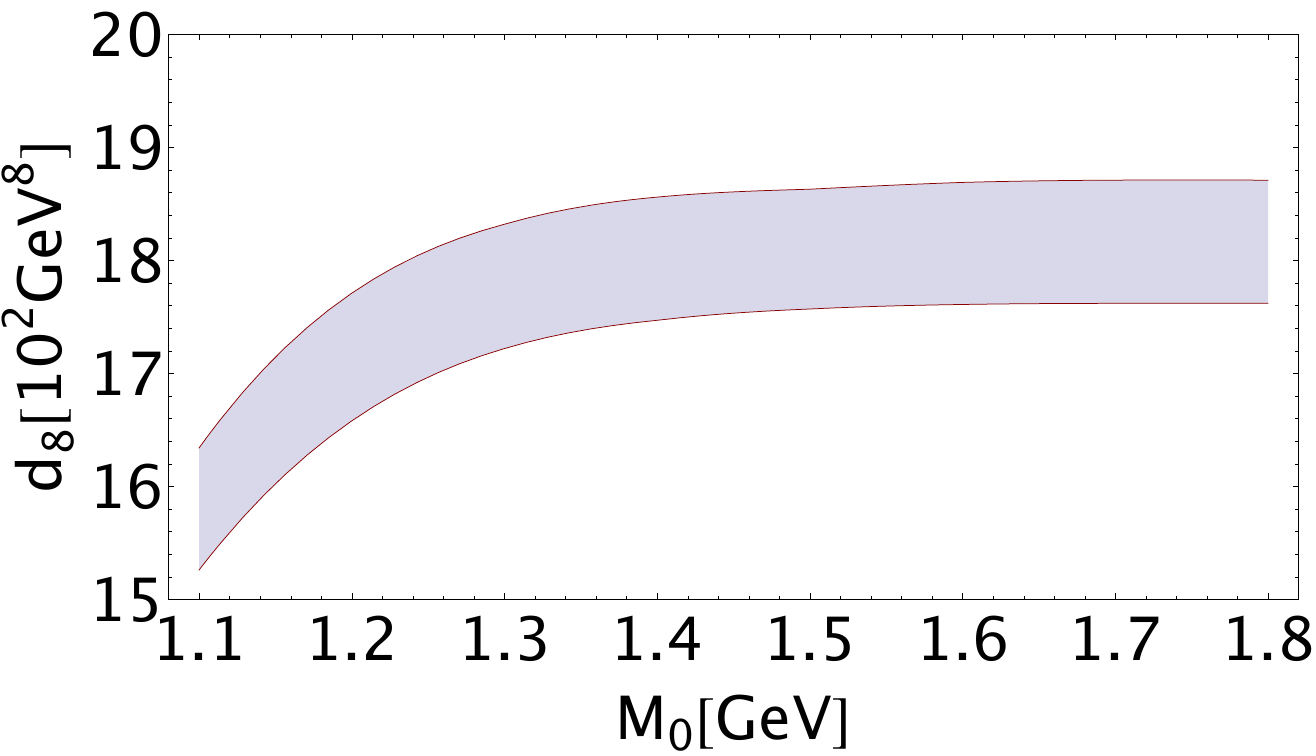}
\includegraphics[width=7.5cm]{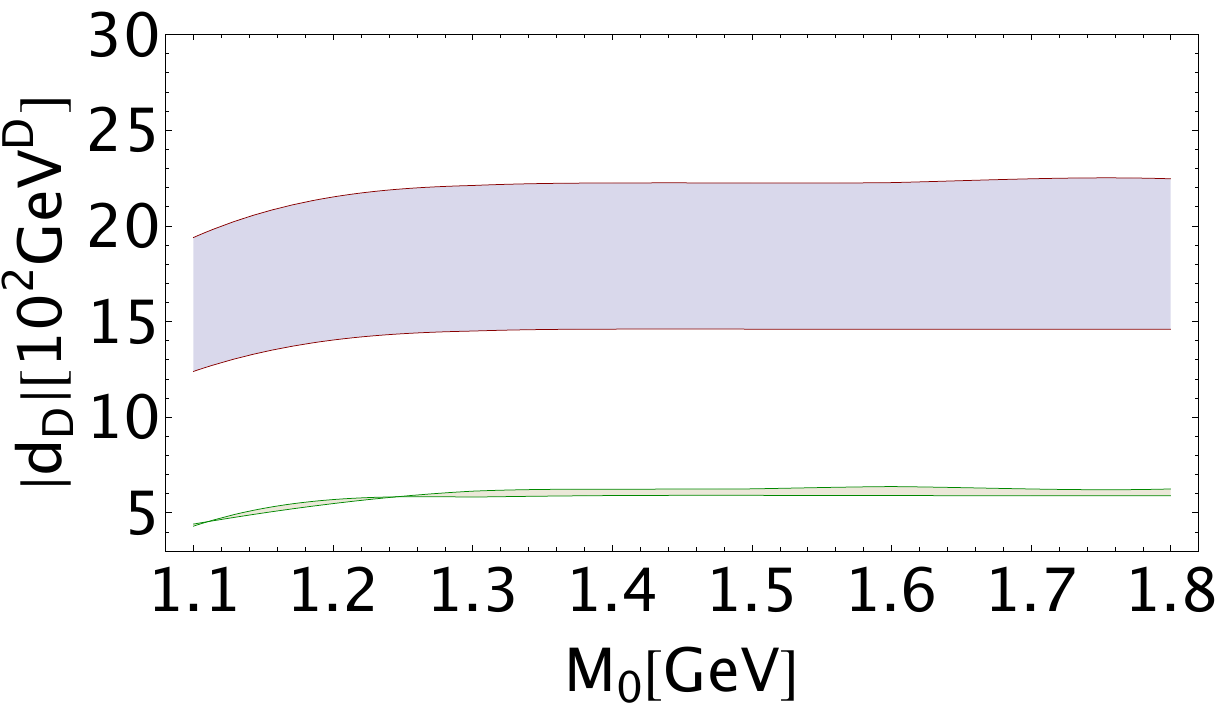}
\caption{\footnotesize {\bf a)} Value of the condensate $d_8$ from ${\cal R}_1^{ee}$ versus $M_0$;
{\bf b)} Values of the condensates $d_{10}$ and $d_{12}$ from ${\cal R}_1^{ee}$ and 
${\cal R}_2^{ee}$  versus $M_0$}\label{fig:d8}
\end{center}
\vspace*{-0.5cm}
\end{figure}

\begin{figure}[hbt]
\begin{center}
\hspace*{-4cm}{\bf a)}\hspace*{8cm} {\bf b)}\\
\includegraphics[width=7.5cm]{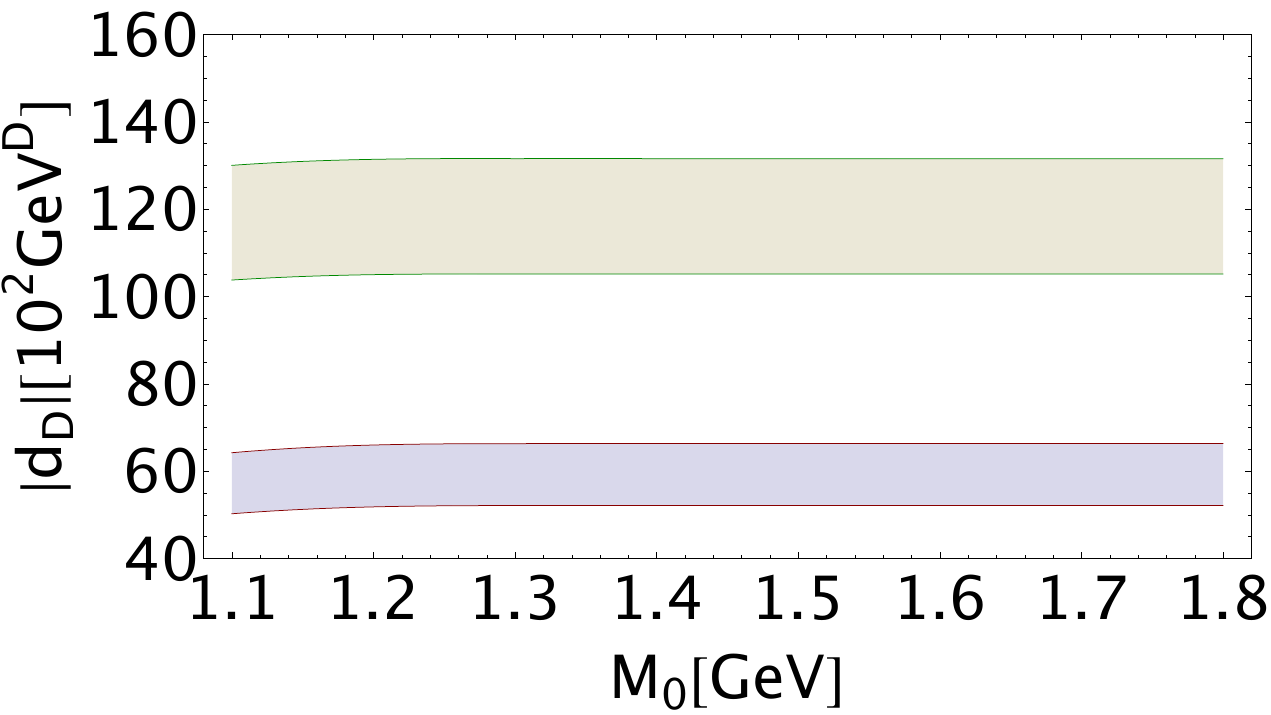}
\includegraphics[width=7.5cm]{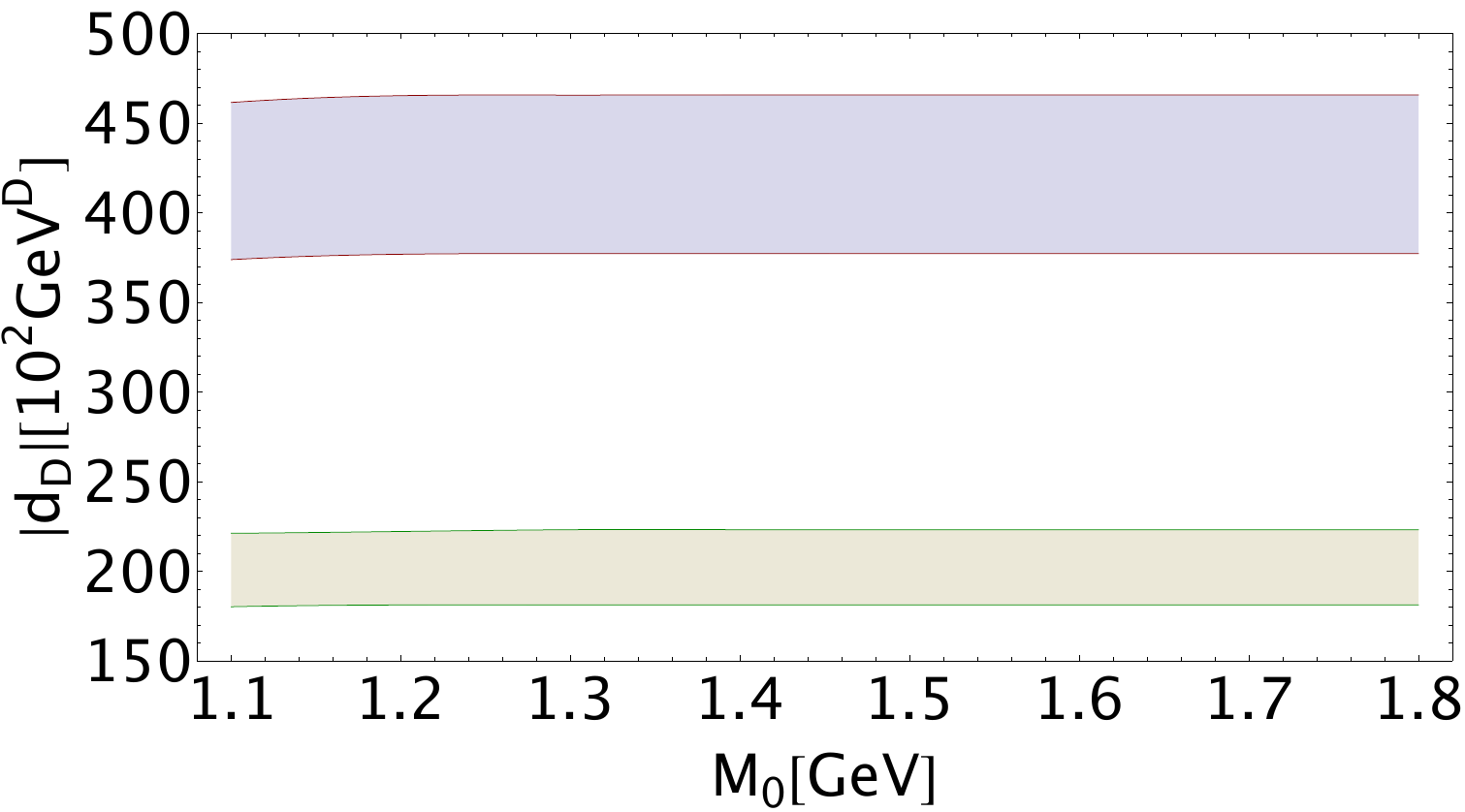}
\caption{\footnotesize {\bf a)} Value of the condensate $d_{14}$ and $d_{16}$ from ${\cal R}_3^{ee}$
and ${\cal R}_4^{ee}$  versus $M_0$;
{\bf b)} Value of the condensates $d_{18}$ and $d_{20}$  from ${\cal R}_5^{ee}$ and 
${\cal R}_6^{ee}$  versus $M_0$}\label{fig:d14}
\end{center}
\vspace*{-0.5cm}
\end{figure}

%%%%%%%%%%%%%%%%%%%%%%%%%
The values of $d_8$ to $d_{20}$ from the $\tau$-like moments are shown in Figs.\,\ref{fig:d8} and \ref{fig:d14} where we notice nice stabilities of the results versus $M_0$.  The optimal values are quoted in Table\,\ref{tab:cond}.
%\end{document}
  %%%%%%%%%%%%%%%%%%%%%%%%%%%%%%%%%%%%%%%
%  \vspace*{-0.5cm}
%%%%%%%%%%%%%%%%%%%%%%%%%%%%%%%%%%%%%%%%%%

 %%%%%%%%%%%%%%%%%%%%%%%%%%%%%%%%%%%%%%%
%  \vspace*{-0.5cm}
%%%%%%%%%%%%%%%%%%%%%%%%%%%%%%%%%%%%%%%%%%

%%%%%%%%%%%%%%%%%%%%%%%%%%%%
\subsection*{\b $d_4$ and $d_6$ from the ratio of LSR}
\begin{figure}[hbt]
\begin{center}
\hspace*{-4cm}{\bf a)}\hspace*{8cm} {\bf b)}\\
\includegraphics[width=7.5cm]{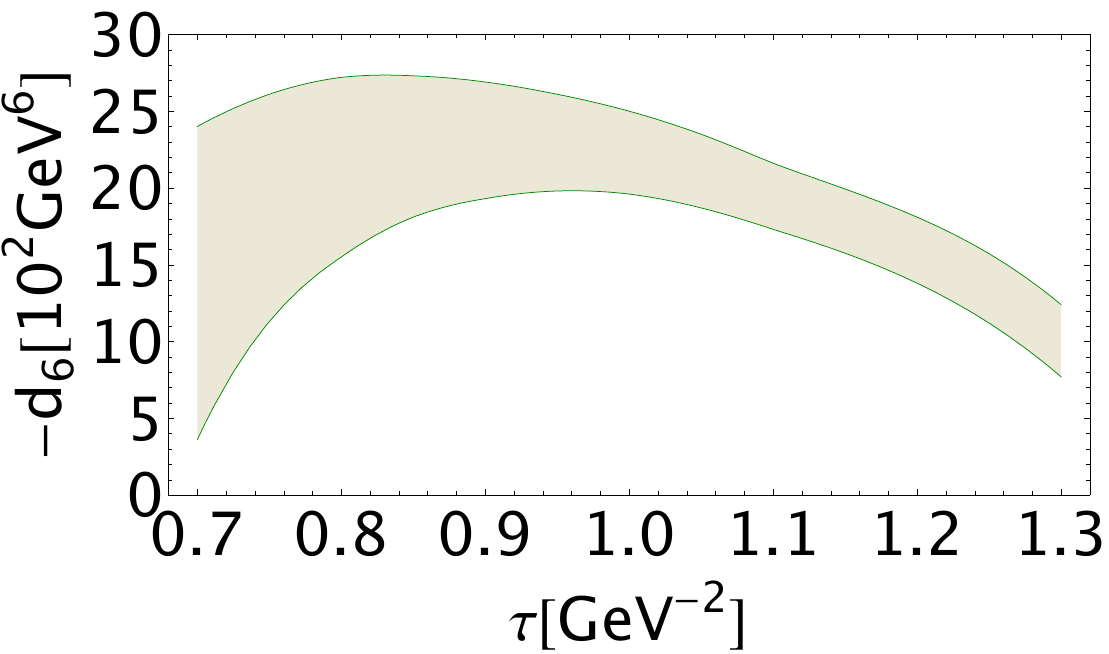}
\includegraphics[width=7.5cm]{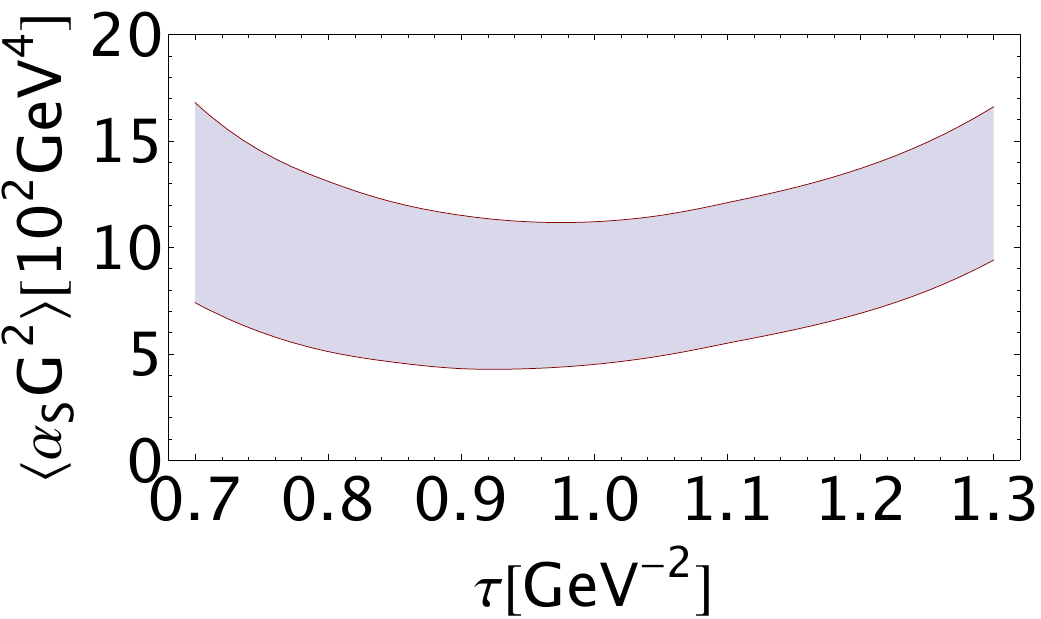}
\caption{\footnotesize {\bf a)} Value of the condensate $d_6$  from ${\cal R}_{10}^{ee}$
versus $M_0$;
{\bf b)} Value of the condensate $d_{4}$   from ${\cal R}_{10}^{ee}$   versus $M_0$}\label{fig:d6}
\end{center}
\vspace*{-0.5cm}
\end{figure}

%%%%%%%%%%%%%%%%%%%%%%%%%%%%
Once we obtain the previous values of the $d=8$ to $d=20$ condensates, we inject them to the ratio ${\cal R}_{10}$ of LSR for re-extracting $d_6$ and $d_4$. To strongly constrain the parameters,
we continue to use a one-parameter fit. Fixing again $\la \alpha_s G^2\ra$ as in Eq.\,\ref{eq:asg2}, we re-extract $d_6$ by including into the QCD expression of ${\cal R}_{10}$ the contributions of condensates up to $d_{20}$. The analysis is shown  in Fig.\,\ref{fig:d6}.
 %%%%%%%%%%%%%%%%%%%%%%%%%%%%%%%%%%%%%%%
%  \vspace*{-0.5cm}
%%%%%%%%%%%%%%%%%%%%%%%%%%%%%%%%%%%%%%%%%%
The results  are summarized in Table\,\ref{tab:cond} and compared with the other estimates
quoted in Table\,\ref{tab:other}. One can notice a good agreement with some published results but the values obtained in the present work are more accurate.
    %%%%%%%%%%%%%%%%%%%%%%%%%%%%%%%%%%%%%%%%
   {\normalsize
\begin{table}[H]
%\tbl{
%}
\setlength{\tabcolsep}{0.22pc}
  \begin{center}
    {%\small
  \begin{tabular}{lllll ll ll}

% {\begin{tabular}{@{}lll@{}} \toprule
&\\
\hline
\hline
%\multicolumn{7}{c}{\bf Our Work}&\multicolumn{3}{c}{Ref.\,\cite{ZHUT}}\\
%\hline
$\oliva \la\alpha_s G^2\ra$&$\oliva -d_6$&$\oliva d_8$&-$\oliva d_{10}$&-$\oliva d_{12}$&$\oliva d_{14}$& $\oliva -d_{16}$&$\oliva -d_{18}$&$\oliva d_{20}$\\
 \hline 
 %%%%%%%%%%%%%
$7.8\pm 3.5$&$23.6\pm 3.7$&$18.2\pm 0.6$&$6.1\pm 0.2$&$18.4\pm 3.8$&$59.2\pm 7.1$&$118.3\pm 13.2$&$ 202.0\pm 20.7$&$421.3\pm 44.3$\\
   \hline\hline
 % \vspace*{-0.75cm}
\end{tabular}}
 \caption{ Values of the QCD condensates of dimension $D$ in units of $10^{-2}$ GeV$^{D}$ from  the present analysis.}\label{tab:cond} 
%}
%\caption{%\scriptsize   
 \end{center}
\end{table}
} 
 %%%%%%%%%%%%%%%%%%%%%%%%%%%%%%%%%
 \vspace*{-1cm}
    %%%%%%%%%%%%%%%%%%%%%%%%%%%%%%%%%%%%%%%%
   {\scriptsize
   \begin{center}
\begin{table}[hbt]
%\tbl{
%}
\setlength{\tabcolsep}{0.6pc}
  \begin{center}
    {%\small
  \begin{tabular}{lllll ll}

% {\begin{tabular}{@{}lll@{}} \toprule
&\\
\hline
\hline
%\multicolumn{7}{c}{\bf Our Work}&\multicolumn{3}{c}{Ref.\,\cite{ZHUT}}\\
%\hline
\oliva$\la\alpha_s G^2\ra$ &\oliva$-d_6$ &\oliva$d_8$&\oliva$-d_{10}$&\oliva$-d_{12}$  &\oliva Refs.\\
 \hline 
 %%%%%%%%%%%%%
$0.67\pm 0.89$&$15.2\pm 2.2$&$22.3\pm 2.5$&&& {\rm ALEPH\,\cite{ALEPH}}\,\\
$5.34\pm 3.64$&$14.2\pm 3.5$&$21.3\pm 2.5$&&& {\rm OPAL\,\cite{OPAL}}\,\\
 $3.5^{+2.2}_{-3.8}$&$19.7^{+11.8}_{-7.9}$&$23.7^{+11.8}_{-15.8}$&$11.8\pm 19.7$&$7.9\pm 19.7$&(${  d_{14,16}=0}$){  Pich-Rodriguez\,\cite{PICH1}}\\
 % &$19.7\pm 9.9$&$23.7\pm 16.6$&$11.8\pm 19.7$&$7.9\pm 19.7$&{\tiny Pich-Rodriguez}\\

%$0.31\pm 2.45$&$13.5\pm 1.8$&$20.0\pm 1.6$&       {\rm CI}&\tiny {\rm OPAL}\\\
%$-1.57\pm0.94$&$14.7\pm 1.1$&$20.4\pm 1.3$&{\rm CI}&\tiny{\rm ALEPH}\\
%$-0.8^{+0.7}_{-0.7}$&$35\pm 3$&$49 ^{+4}_{-5}$&  {\rm CI}&\tiny PICH-RODRIGUEZ\\
   \hline\hline
 % \vspace*{-0.75cm}
\end{tabular}}
 \caption{Values of the QCD condensates from some other $\tau$-moments at Fixed Order (FO) of the PT series.}\label{tab:other} 
 \end{center}
\end{table}
\end{center}
} 
 %%%%%%%%%%%%%%%%%%%%%%%%%%%%%%%%%
 \subsection*{\b Comments on the results}
  %%%%%%%%%%%%%%%%%%%%%%%%%%%%%%%%%
 \d The value of $\la \alpha_s G^2\ra$ obtained here is compatible with the one in Eq.\,\ref{eq:asg2} but less accurate. However, we consider this result as an improvement of the ones from some other $\tau$-decay moments given in Table\,\ref{tab:other} using FO perturbative series. It disfavours some negative values obtained from CI perturbative series. This value is also consistent with the one $(8\pm 4)\times 10^{-2}$ GeV$^4$ obtained from a matching of the short and long distances contributions to the Adler function at large $N_c$\,\cite{PERROTET}. 
 
 \d  The value of $d_6$ is in fair agreement with the one in Eq.\,\ref{eq:d6} from ratio of LSR obtained in SN23\,\cite{SNe} where the OPE has been truncated at $D=8$. The inclusion of higher dimension condensates has increased slightly $|d_6|$ by about 13.6\%. The ratio 
 $\la \alpha_s G^2\ra/\rho\alpha_s\la\bar\psi\psi\ra^2=95.4$ is in the range  of the ones obtained using different methods quoted in\,\cite{SN95}. 
 
 \d The value of $d_8$ is larger than the one in Eq.\,\ref{eq:d8} from ratio of LSR. As the extraction of $d_8$ from LSR in SN23 has been done by neglecting all higher dimension condensates included in the present analysis, we interpret the value of $d_8$ obtained in SN23 as an effective condensate including all high dimension ones. To check this argument, we evaluate  the NP contributions to ${\cal R}_{10}^{ee}$ at the typical  optimization sum rule scale $\tau\approx 1 $  GeV$^{-2}$.  We neglect all higher dimension condensates and find\,:
 \beq
 \sum_4^8 \delta^{NP}= 5.1\,\times 10^{-2},~~~~\sum_4^6 \delta^{NP}= -1.0\,\times 10^{-2} \rar d_8\approx 6.1\,\times 10^{-2}\,{\rm GeV}^{8}
 \eeq
 in perfect agreement with the fitted value  in Eq.\,\ref{eq:d8}.
 
\d Using the previous values of the condensates, we check the convergence of the OPE. We obtain at $\tau\simeq 1$ GeV$^{-2}$:
\beq
 \sum_{10}^{20} \delta^{NP}= -0.4\,\times 10^{-2},
 \eeq
 which is negligible. This feature justifies (a posteriori) the phenomenological success of the SVZ sum  rules by truncating the OPE up to dimension $D=6,8$ condensates.
 
 \d From Table\,\ref{tab:cond}, we observe that there is no signal of a factorial nor/and exponential growth of the size of the condensates up to $D=20 $ where the OPE is truncated. Similar observation can be made in the $V-A$ channel for the OPE up to $D=18$\,\cite{SNV-A,PICHV-A}. 
  %%%%%%%%%%%%%%%%%%%%%%%%%%%%%%%%%
 \subsection*{\b Comments on the ${\cal L}_0$ LSR moment}
  %%%%%%%%%%%%%%%%%%%%%%%%%%%%%%%%%
We have also tried to use the ${\cal L}_0$ LSR moment to fix the $D\leq 8$ dimension condensates\,\footnote{Some other LSR moments with weights are discussed in\,\cite{SNV-A,CVETIC}.}.  Our analysis is unconclusive and confirms the observation done in SN23. 
The same feature is observed for an attempt to extract $\alpha_s$. 
 %%%%%%%%%%%%%%%%%%%%%%%%%%%%%%%%%%%%%%%%
\section{$\alpha_s$ from the lowest BNP moment using $e^+e^-\to$ Hadrons data}
%%%%%%%%%%%%%%%%%%%%%%%%%%%%%%%%%%%%%%%%%
Armed with these improved values of the QCD condensates, we  extract $\alpha_s$ from the lowest BNP moment ${\cal R}^{ee}_0$. For convenience, we rescale at $M_\tau$ the value of $\alpha_s$ obtained for different values of $M_0$ 
 %%%%%%%%%%%%%%%%%%%%%%%%%%%%%%%%%
 \subsection*{\b Standard SVZ expansion}
  %%%%%%%%%%%%%%%%%%%%%%%%%%%%%%%%%
 \subsection*{\hspace*{0.5cm} \d Fit of the data points}
\begin{figure}[hbt]
\begin{center}
%\hspace*{-4cm}{\bf a)}\hspace*{8cm} {\bf b)}\\
\includegraphics[width=14cm]{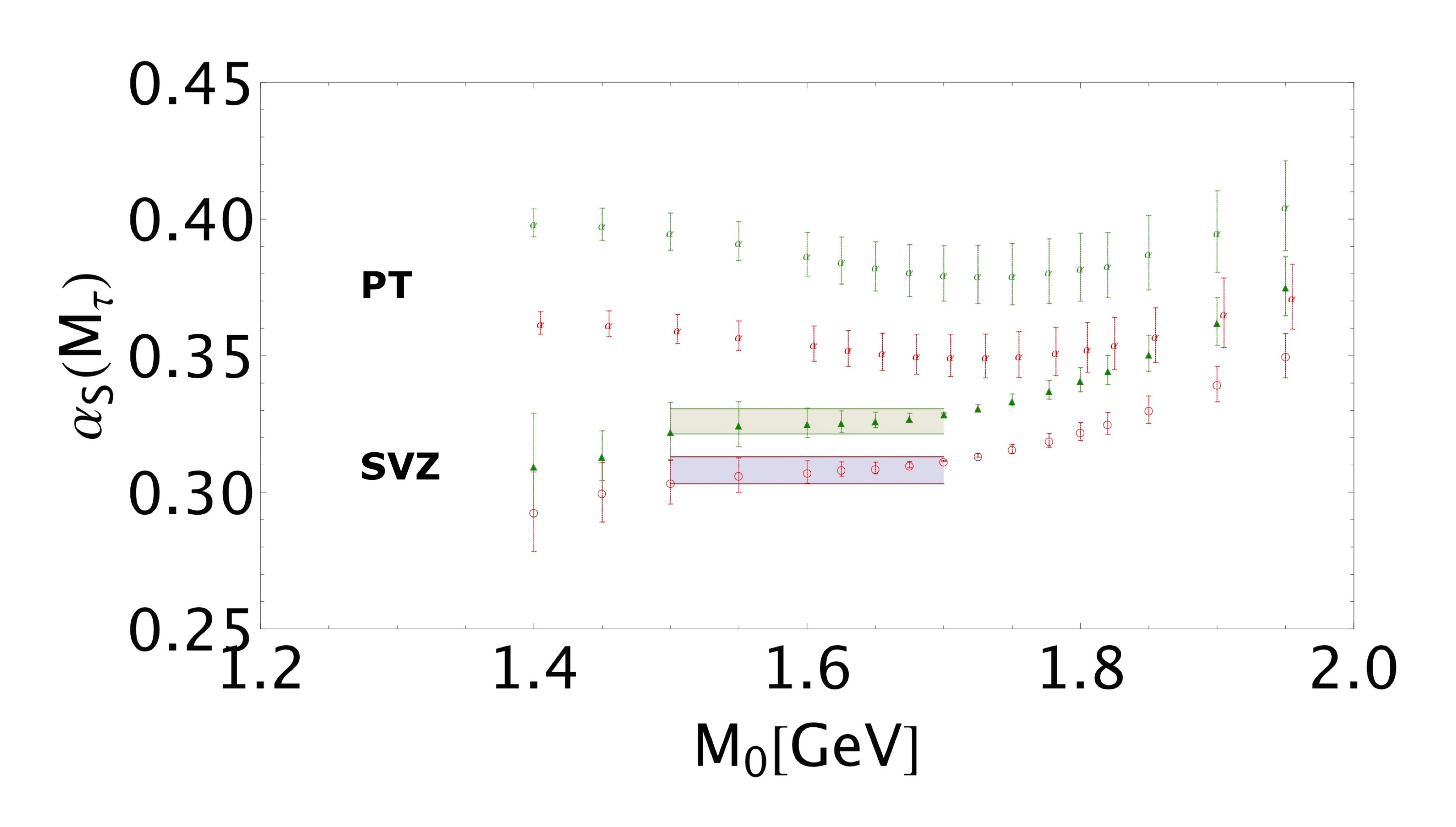}
\caption{\footnotesize {\bf a)} Value of $\alpha_s (M_\tau)$ from the BNP lowest moment
for different values of $M_0$. The red (green) points correspond to (FO) [resp. (CI)] PT series.
The lower curves correspond to the SVZ expansion. The upper ones to the case of zero value of the QCD condensates.
}\label{fig:as}
\end{center}
\vspace*{-0.5cm}
\end{figure}

 %%%%%%%%%%%%%%%%%%%%%%%%%%%%%%%%%
The analysis is shown in Fig.\,\ref{fig:as} (lower set of points) for (FO) and (CI) perturbative series to order $\alpha_s^4$. We observe  a stability for $M_0$  between 1.5 to 1.7 GeV. 
A least square fit of the data points gives:
\bea
\alpha_s(M_\tau) &=& 0.3081(49)_{fit}~~~~~~{\rm (FO)}\nnb\\
&=& 0.3260(47)_{fit} ~~~~~~{\rm (CI)}
\eea
  %%%%%%%%%%%%%%%%%%%%%%%%%%%%%%%%%
 \vspace*{-1cm}
%%%%%%%%%%%%%%%%%%%%%%%%%%%%%%%%%%%%%%%%%%

  %%%%%%%%%%%%%%%%%%%%%%%%%%%%%%%%%
 \subsection*{\hspace*{0.5cm} \d Estimate of the $\alpha_s^5$ corrections}
  %%%%%%%%%%%%%%%%%%%%%%%%%%%%%%%%%
  Noting that the coefficient of $\alpha_s$ for the Adler $D$-function grows geometrically\,\cite{SNZ}:
  \beq
 D(Q^2)=\sum_n{a_s^n c_n}~: ~c_0=c_1=1,~~c_2=1.656,~c_3=6.37,~c_4=49.09  
 \eeq
 we deduce:
 \beq
  c_4\approx c_3^2~\lrar2 ~c_5\simeq\ga {c_3}/{c_2}\dr c_4^2\simeq 228. 
 \eeq
 This value of $c_5$ is comparable with some other estimates from FAC\,: $\simeq 275$\,\cite{CHET4}, conformal mappings\,: $=255$\,\cite{CAPRINI},  Pad\'e approximants\,: $=(277\pm 51)$\cite{MASJUAN} and a linear behavior of FO coefficients\,: $\simeq 283$\,\cite{BENEKE}. Writing the QCD expression of the moment as:
 \beq
 {\cal R}^{ee}_0=\sum_n a_s^n (g_n+c_n)~,
 \eeq
 where $g_n$ comes from a  RG-resummation\,\cite{LEDI,KATAEV}. One obtains\,:
 \bea
 g_5&=&-780  ~~~ \lrar2~~~ \Delta \alpha^5_s(M_\tau) \simeq \pm 71\times 10^{-4} ~~~{\rm (FO)}, \nnb\\
 &=& 0~~~~~~ ~~ \lrar2~~~  \Delta \alpha^5_s(M_\tau) \simeq \pm 62\times 10^{-4}~~~~{\rm (CI)}
 \vspace*{-0.5cm}
 \eea
  %%%%%%%%%%%%%%%%%%%%%%%%%%%%%%%%%
 \subsection*{\hspace*{0.5cm} \d Final value of $\alpha_s$ using the SVZ expansion}
  %%%%%%%%%%%%%%%%%%%%%%%%%%%%%%%%%
Adding the different corrections, we deduce:
\bea
\alpha_s(M_\tau)\vert_{e^+e^-}^{SVZ}  &=& 0.3081(49)_{fit}(71)_{\alpha_s^5} ~~~\lrar2~~~\alpha_s(M_Z)\vert_{e^+e^-}^{SVZ}  = 0.1170(6)(3)_{evol} ~~~~~~   {\rm (FO)}\nnb\\
&=& 0. 3260(47)_{fit}(62)_{\alpha_s^5} ~~~\lrar2~~~\alpha_s(M_Z)\vert_{e^+e^-}^{SVZ}  = 0.1192(6)(3)_{evol} ~~~~~~{\rm (CI)},
\label{eq:as-foci}
\eea
where the sum of non-perturbative corrections at the $\tau$-mass is given in Table\,\ref{tab:tau}.
 We notice that the  increase of $|d_6|$ has slightly decreased the value  of $\alpha_s$
 compared to the one in \cite{SNe}. 
From the previous values, we attempt to estimate the mean:
\beq
\alpha_s(M_\tau)\vert_{e^+e^-}^{SVZ} = 0.3179(58)(81)_{syst} ~~~\lrar2~~~\alpha_s(M_Z)\vert_{e^+e^-}^{SVZ}  = 0.1182(12)(3)_{evol}.
\label{eq:as-e+e-}
\eeq
 %%%%%%%%%%%%%%%%%%%%%%%%%%%%%%%%%
 \subsection*{\b QCD models without power corrections}
%%%%%%%%%%%%%%%%%%%%%%%%%%%%%%%%
 Here, we analyze a QCD model without power corrections. We show the result in Fig\,\ref{fig:as} (upper set of curves). We obtain at the minimum $M_0\simeq 1.75$ GeV of the curves:
\bea
\alpha_s(M_\tau)\vert_{e^+e^-}^{PT}   &=& 0.3579(115)_{fit}(34)_{\alpha_s^5} ~~~\lrar2~~~\alpha_s(M_Z)\vert_{e^+e^-}^{PT}  = 0.1228(12)(3)_{evol} ~~~~~~   {\rm (FO)}\nnb\\
&=& 0. 3855(117)_{fit}(83)_{\alpha_s^5} ~~~\lrar2~~~\alpha_s(M_Z)\vert_{e^+e^-}^{PT}  = 0.1255(13)(3)_{evol} ~~~~~~{\rm (CI)},
\eea
which corresponds to the mean:
\beq
\alpha_s(M_\tau)\vert_{e^+e^-}^{PT}  = 0.3692(92)(163)_{syst} ~~~\lrar2~~~\alpha_s(M_Z)\vert_{e^+e^-}^{PT}  = 0.1239(18)(3)_{evol}.
\eeq
This result is about 4.6 $\sigma$  higher than the PD23 weighted average\,\cite{PDG}:
\beq
\la\alpha_s(M_Z)\ra= 0.1178(5),
\label{eq:pdg}
\eeq
 and disfavors some QCD models without power corrections. 
%%%%%%%%%%%%%%%%%%%%%%%%%%%%%%%%%%%%%%%%%%%%
\section*{9. Comparison of $\alpha _{s}$ from $e^+e^-$ and $\tau $-decay data within the SVZ expansion.}
\label{sec9}
%%%%%%%%%%%%%%%%%%%%%%%%%%%%%%%%%%%%%%%%%
%%%%%%%%%%%%%%%%%%%%%%%%%%%%%%%%%%%%%%%%%%%%%%%%%%%%%%%%%%%%%%%%%%%%%%%%%%%%%%%%%%%%%%%%%%
%%%%%%%%%%%%%%%%%%%%%%%%%%%%%%%%%%%%%%%%%%%%%
\subsection*{\b Update of $\alpha_s$ from  $\tau$-decay using the lowest BNP moment}
%%%%%%%%%%%%%%%%%%%%%%%%%%%%%%%%%%%%%%%%%%%%%%
Within the new set of non-perturbative contributions, we update the determination of $\alpha_s$ from $\tau$-decay. We shall consider the Vector (V) channel. The QCD expression of the lowest moment reads\,\cite{BNP}:
\beq
{\cal R}^{qcd}_{\tau,V}=\frac{3}{2}\,|V_{ud}|^2 S_{ew}\Bigg{[}1+\delta'_{ew}+ \delta^{(0)}_0+\delta_{ud}^{(D=4)}-6\frac{d_6}{M_\tau^6}-4\frac{d_8}{M_\tau^8}\Bigg{]},
\eeq
where  $\delta_0^{(0)}$ is the PT corrections and $\delta_{ud}^{(D=4)}$ is the dimension 4 contributions given in Eq. 3.11 of\,\cite{BNP}. The values of $d_6$ and $d_8$ condensate contributions are given inTable\,\ref{tab:cond}, while the electroweak parameters are\,\cite{BNP}:
\beq
|V_{ud}|=0.97425(22),~~~~~~~~~~~~~~~~S_{ew}=1.0194,~~~~~~~~~~~~~~~~\delta'_{ew}=0.0010.
\eeq
Using the updated ALEPH data \,\cite{DAVIER}:
\beq
{\cal R}^{exp}_{\tau,V}= 1.782\pm 0.009,  
\eeq
we obtain from the lowest moment: 
\bea
\alpha_s(M_\tau)\vert_{\tau,V} &=& 0.3128(19)_{fit}(77)_{\alpha_s^5} ~~~\lrar2~~~\alpha_s(M_Z)\vert_{\tau,V} = 0.1176(10)(3)_{evol} ~~~~~~   {\rm (FO)}\nnb\\
&=& 0. 3291(25)_{fit}(65)_{\alpha_s^5} ~~~\lrar2~~~\alpha_s(M_Z) \vert_{\tau,V}= 0.1196(8)(3)_{evol} ~~~~~~{\rm (CI)},
\label{eq:as-foci-tau}
\eea
which corresponds to the mean:
\beq
\alpha_s(M_\tau)\vert_{\tau,V} = 0.3219(52)(91)_{syst} ~~~\lrar2~~~\alpha_s(M_Z)\vert_{\tau,V}= 0.1187(13)(3)_{evol}.
\label{eq:tau}
\eeq

%%%%%%%%%%%%%%%%%%%%%%%%%%%%%%%%%%%%%%%%%%%%%%%%
\subsection*{\b Average value of $\alpha_s$ from  $e^+e^-$ and $\tau$-decay data}
%%%%%%%%%%%%%%%%%%%%%%%%%%%%%%%%%%%%%%%%%%%%%%%%%%
One can notice an excellent agreement between the value of $\alpha_s(M_\tau)$ from $e^+e^-$ in Eq.\,\ref{eq:as-e+e-} and the one from the vector component of $\tau$-decay in Eq.\,\ref{eq:tau}.  We consider as a final value of $\alpha_s$ from the present analysis based on the lowest $\tau$-decay-like moment,
the average of the results in Eqs.\,\ref{eq:as-e+e-} and \ref{eq:tau} which is:
\beq
\la \alpha_s(M_\tau)\ra = 0.3198(72) ~~~\lrar2~~~\la\alpha_s(M_Z)\ra= 0.1185(9)(3)_{evol}.
\label{eq:as-final}
\eeq
This mean value is in excellent agreement with the latest  PDG average given in Eq.\,\ref{eq:pdg}. 
%%%%%%%%%%%%%%%%%%%%%%%%%%%%%%%%%%%%%%%%%%%%%%%%%%%%%%%%%%%%%%%%%%%%%
\subsection*{\b Comparison with $\alpha_s$ from  $\tau$-decay data within the SVZ expansion in the literature}
%%%%%%%%%%%%%%%%%%%%%%%%%%%%%%%%%%%%%%%%%%%%%%%%%%%%%%%%%%%%%%%%%%%%%

%%%%%%%%%%%%%%%%%%%%%%%%
Estimates of $\alpha _{s}$ using the vector component of $\tau $-decay
data within different $\tau $-decay moments are available in the literature\,\cite{ALEPH,OPAL,PICH1,DAVIER}
as shown in Table ~\ref{tab:tau}. One can notice that the central value
of $\alpha _{s}(M_{\tau})$ from $e^{+}e^{-}\to $ Hadrons data using the
lowest BNP moment is slightly lower than the one from $\tau $-decay data
but agrees within the errors. We also note that our value of
$\delta ^{(NP)}$ at the $\tau $-mass agrees with the OPAL one but higher
than the ones using ALEPH data.
 %%%%%%%%%%%%%%%%%%%%%%%%%%%%%%%%%
% \vspace*{-1cm}
    %%%%%%%%%%%%%%%%%%%%%%%%%%%%%%%%%%%%%%%%

  {\normalsize
     \begin{center}
\begin{table*}[hbt]
% space before first and after last column: 1.5pc
% space between columns: 3.0pc (twice the above)
\setlength{\tabcolsep}{0.2pc}
% -----------------------------------------------------
% adapted from TeX book, p. 241
\newlength{\digitwidth} \settowidth{\digitwidth}{\rm 0}
\catcode`?=\active \def?{\kern\digitwidth}
% -----------------------------------------------------
%\caption{Biologically treated effluents (mg/l)}
\label{tab:effluents}
\begin{tabular*}{\textwidth}{@{}l@{\extracolsep{\fill}}cccccc c}
\hline
              & \multicolumn{2}{c}{\oliva THIS WORK} 
                  &\oliva ALEPH\,\cite{ALEPH}&\oliva OPAL\,\cite{OPAL} &\oliva PR\,\cite{PICH1}&\oliva ALEPH\,\cite{DAVIER}\\
                          % & \multicolumn{2}{l}{Full scale plant} \\
\cline{2-3} %\cline{4-5}
                 & \multicolumn{1}{c}{$e^+e^-$} 
                 & \multicolumn{1}{c}{$\tau$-decay} \\
                   \hline

FO&0.3081(86)&0.3128(79)& 0.3200(220)&0.3230(160)&0.3200(150)&--&\\
CI & 0.3260(78)&0.3291(70)&0.3400(230)&0.3470(220)&0.3370(200)&0.3460(110) \\
FO $\oplus$ CI&0.3179(100)&0.3219(105) \\
$\delta^{(NP)}\,\times 10^{-2}$&$3.8\pm 0.8$&$3.7\pm1.0$&$2.0\pm 0.3$&$3.6\pm 0.4$&$1.7\pm 0.3$&$2.0\pm 0.3$&\\
\hline
\vspace*{-0.5cm}
\end{tabular*}
 \caption {\footnotesize  $\alpha_s(M_\tau)$ within the SVZ-expansion from the vector (V) component of $\tau$-decay data using some high-$\tau$-moments. For a better comparison with our result where the condensates have been estimated within (FO), we quote the value of $\delta^{(NP)}$ corresponding to (FO) at $M_\tau$. We also show our prediction for the V component of $\tau$-decay using $d_6$ and $d_8$ in Table\,\ref{tab:cond}.}\label{tab:tau}
\end{table*}
 \end{center}
}
\vspace*{-0.5cm}
%%%%%%%%%%%%%%%%%%%%%%%%%%%%%%%%%%%%%%%%%
\section{The quest of $1/Q^2$ and tachyonic gluon mass beyond the SVZ expansion}
%%%%%%%%%%%%%%%%%%%%%%%%%%%%%%%%%%%%%%%%%

\subsection*{\b The quest of $1/Q^2$}

A priori, this question is irrelevant  as one does  not expect to have a dimension-two condensate $\la A^2\ra$ ($A_\mu$is the gluon field) due to gauge invariance (see however\,\cite{STOD}).  However, this term is present in some holographic models\,\cite{HOLO} and in a lattice calculation of the QCD potential\,\cite{BALI} and gluon condensates\,\cite{RAKOW1,RAKOW2}. It  has been also proposed\,\cite{ALT,ZAK} for a phenomenological parametrization of UV renormalon as an alternative to the large $\beta$-approximation. The large $\beta$-approximation is not yet fully justified due to  the non-observation of the factorial growth and alternate signs of the calculated coefficients of the Adler $D$-function and some other QCD observables known to order $\alpha_s^4$. Instead, these calculated coefficients grow as a geometric sum\,\cite{SNZ}.

%%%%%%%%%%%%%%%%%%%%%%%%%%
%\vspace*{-0.25cm}
\subsection*{\b Tachyonic gluon mass $\lambda^2$}

\begin{figure}[hbt]
\begin{center}
\includegraphics[width=8.cm]{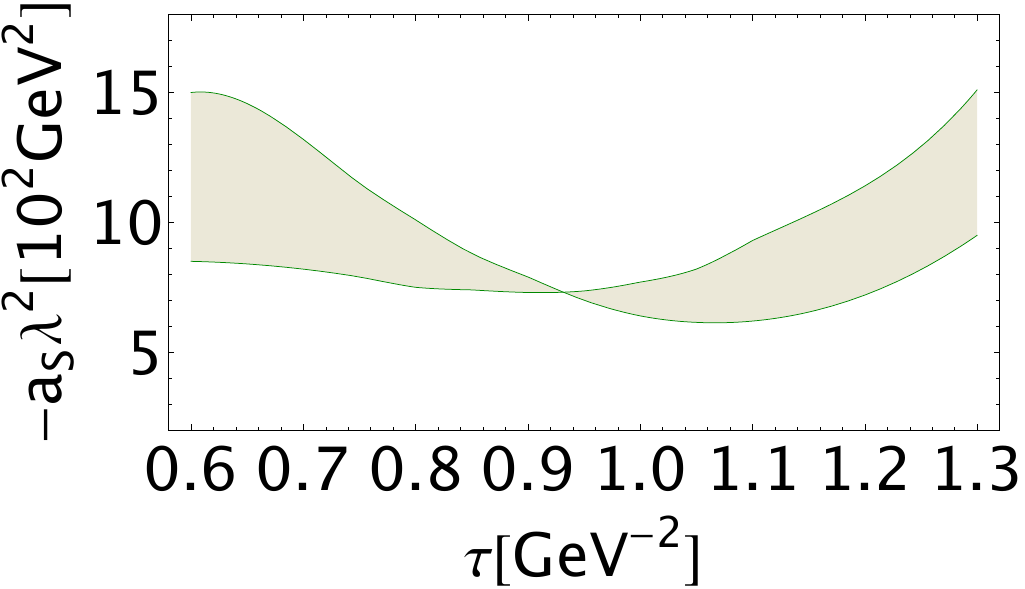}
\caption{\footnotesize The tachyonic gluon mass squared  $a_s\lambda^2$ versus $M_0$.}\label{fig:d2}
\end{center}
\vspace*{-0.5cm}
\end{figure}

%\vspace*{-0.25cm}
%%%%%%%%%%%%%%%%%%%%%%%%%%
\hspace*{0.5cm} \d  In\,\cite{CNZ,D2rev}, a systematic way to account for a such $1/Q^2$ term is the introduction of a tachyonic gluon mass squared $\lambda^2$. An estimate of this mass using  $e^+e^-\to$ Hadrons data and some other data lead to,\cite{SN93,SN95,CNZ,TERAYEV}\,:
 \beq
a_s \lambda^2= -(7\pm 3)\times 10^{-2}~{\rm GeV}^2,
\eeq
which we re-estimate here from the ratio of LSR ${\cal R}_{10}^{ee}$. The result is shown in Fig.\,\ref{fig:d2} from which we deduce  to order $\alpha_s^4$ at the stability regions\,:
\beq
a_s \lambda^2= -(6.7\pm 1.5)\times 10^{-2}~{\rm GeV}^2.
\eeq
 %%%%%%%%%%%%%%%%%%%%%%%%%%%%%%%%%%%%%%%
%  \vspace*{-0.5cm}
%%%%%%%%%%%%%%%%%%%%%%%%%%%%%%%%%%%%%%%%%%%
in perfect agreement with the previous result.  

\begin{figure}[hbt]
\begin{center}
\hspace*{-4cm}{\bf a)}\hspace*{8cm} {\bf b)}\\
\includegraphics[width=7.cm]{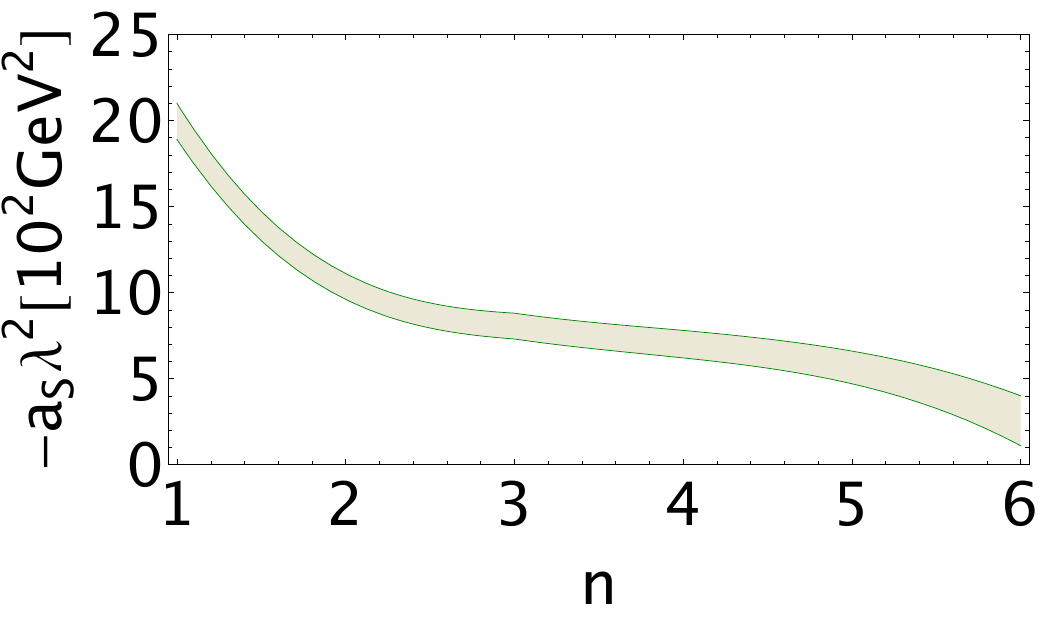}
\includegraphics[width=7.cm]{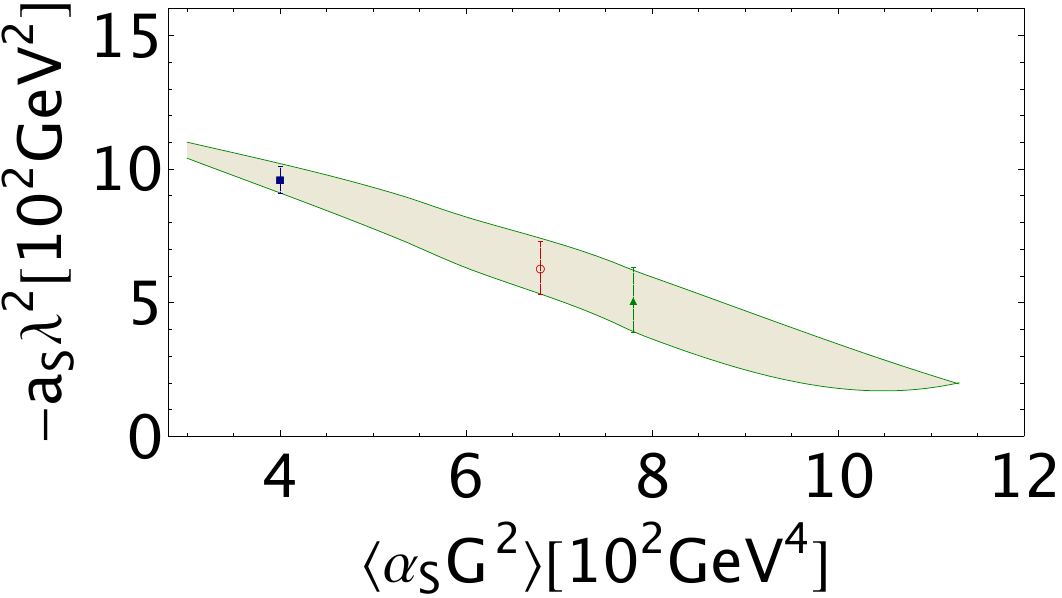}
\caption{\footnotesize {\bf a)} $a_s\lambda^2$  for different truncations of the PT series;
{\bf b)} $a_s\lambda^2$  for different values of the gluon condensate $\la\alpha_s G^2\ra$: the blue point is the SVZ value of $\la\alpha_s G^2\ra$, the red point is the value in Eq.\,\ref{eq:asg2}, the green point is the value quoted in Table\,\ref{tab:cond}. }\label{fig:trunc}
\end{center}
\vspace*{-0.5cm}
\end{figure}

\hspace*{0.5cm}\d A such value of $\lambda^2$  resolves the hierarchy of the $\pi,\rho$ mesons and glueball sum rule scales\,\cite{CNZ}. However, a direct inclusion of this effect to the $\tau$-decay moment is quite delicate.  It has been argued in Ref.\cite{CNZ} that this term is dual to the large order terms of the PT series and decreases when more terms in the PT series are added. To check this argument, we study the variation of $\lambda^2$ versus the number of $\alpha_s$ term added in the PT series of ${\cal R}_{10}$.  We show the analysis in Fig.\,\ref{fig:trunc}a where we, indeed, see that $|\lambda^2|$ decreases when more $\alpha_s^n$ terms are added in the PT series even at low values of $n$. A similar feature is observed at large orders in the calculation of the gluon condensate $\la \alpha_s G^2\ra$ on the lattice\,\cite{RAKOW1,RAKOW2}. 
 %%%%%%%%%%%%%%%%%%%%%%%%%%%%%%%%%%%%%%%
%  \vspace*{-0.5cm}

We have also checked that the change of $d_6$  does not almost affect the size of $\lambda^2$ while we show in Fig.\,\ref{fig:trunc}b the effect of $\la\alpha_s G^2\ra$. 
From the above results, we consider that the $\lambda^2$ effects to the BNP moment is:
\beq
\delta^{(2)}_{\lambda^2} \leq 4\%,
\eeq
of  the PT lowest order one which is much less than the size of sum of the $\alpha_s$ corrections.  

\hspace*{0.5cm}\d Assuming that our estimate of $\alpha_s^5$ correction to the $\tau$-decay moment is a good approximation of the contribution of higher order terms and using the duality property  with $\lambda^2$ advocated by Ref.\,\cite{SNZ}, we do not add this $\lambda^2$ correction in ${\cal R}_0^{ee}$ to avoid a double counting.  
%%%%%%%%%%%%%%%%%%%%%%%%%%%%%%%%%%%%%%%%%
\section{Some other contributions beyond the SVZ expansion}
%%%%%%%%%%%%%%%%%%%%%%%%%%%%%%%%%%%%%%%%%
\subsection*{\b Small size instantons }
%%%%%%%%%%%%%%%%%%%%%%%%%%%%%%%%%%%%%%%%%
%Instanton effects have been studied in Refs.\,\cite{BALITSKY,NASON,KART} where they are expected to manifest at order $1/M_\tau^9$ or $1/M_\tau^{18}$.  

\hspace*{0.5cm}\d Ref.\,\cite{NASON} found that these corrections can be parametrized as:
\beq
\delta^{inst}\simeq \ga\frac{3.64\,\Lambda}{M_\tau}\dr^9\ga\frac{\hat m_u\hat m_d\hat m_s}{M_\tau^3}\dr\simeq 1.6\,\times 10^{-8},
\eeq
where \,\cite{SNmass} $\hat m_u=3.1(2)$ MeV, $\hat m_d=6.1(4)$ MeV and $\hat m_s=(114.3\pm 6.4)$ MeV are the renormalization group invariant light quark masses and we use $\Lambda=0.34$ GeV. This contribution is completely negligible.

\hspace*{0.5cm}\d In Ref.\,\cite{BALITSKY}, the instanton effect is introduced via the constituent quark mass:
\beq
m_{const} =m_{current} - \frac{2}{3}\pi^2\la \bar \psi\psi\ra\rho^2,
\eeq
where $\rho$ is the instanton size. The authors conclude that at the $\tau$-mass, the effect behaves as $1/M_\tau^{18}$ and  is\,:
\beq
\delta^{inst}\simeq 0.03\sim 0.05,
\eeq
which is comparable with the $D=6$ dimension condensates contribution. However, one should notice that this effect depends crucially on the value of the instanton size $\rho$ which is not under a good control. 

\hspace*{0.5cm}\d Ref.\,\cite{KART} has tested the validity of Ref.\,\cite{BALITSKY} prediction by noting that  Ref.\,\cite{BALITSKY} result contributes to  the $V-A$ $\tau$-decay rate as\,:
\beq
\delta_{V-A}^{inst} \simeq 0.09\sim 0.15,
\eeq
while the ALEPH data\,\cite{ALEPH} only allows the range of values\,:
\beq
\delta_{V-A}^{inst}\simeq (1\pm 3)\,\times 10^{-3}
\eeq
Using the (accepted) relation:  
\beq
\delta_{V+A}^{inst}\simeq \frac{\delta_{V-A}^{inst}}{20},
\eeq
one  concludes that :
\beq
\delta_{V}^{inst}\simeq (0.5\pm 1.5)\,\times 10^{-3},
\eeq
which is less than 5\% of the $d_6$ contribution to the ${\cal R}_\tau^{ee}$ decay rate. Moreover, the result of\,\cite{BALITSKY} is much larger than the direct fit of the dimension $D=18$ term  in the OPE quoted in Table\,\ref{tab:cond}. 
%%%%%%%%%%%%%%%%%%%%%%%%%%%%%%%%%%%%%%%%%
\subsection*{\b Duality violation}
%%%%%%%%%%%%%%%%%%%%%%%%%%%%%%%%%%%%%%%%%
\hspace*{0.5cm}\d This effect is maintained by Ref.\,\cite{BOITO}. It corresponds to an additional QCD contribution to the spectral function introduced by hand as:
\beq
 \Delta {\rm  Im}\,\Pi(t)\vert_{DV}\sim e^{-(\delta+\gamma\,t)}\,\sin (\alpha+\beta\,t),
\eeq
 above a certain threshold :  { $t_c\equiv \hat s_0\simeq 1.5$ GeV$^2$} where $\delta,~\gamma,~\alpha,~\beta$ are free  unknown fitted parameters. 
 
\hspace*{0.5cm} \d The eventual presence of a such term in QCD\,\cite{SHIFMAN} and in some toy-model \,\cite{EDUARDO})  has been discussed at large $N_c$ with an infinite number of narrow resonances
% inspired from 't Hooft model in 2-dimensions\,\cite{THOOFT} at large $N_c$ with an infinite number of narrow resonances 
  in order to modelize the data in the  Minkowski space (see also\,\cite{THOOFT}  for QCD in two dimensions). 
   {\it This oscillating term in the Minkowski space is expected to be dual to an asymptotic exponential behavior of power corrections in the Euclidean region\,\cite{SHIFMAN}}. 
  
 %%%%%%%%%%%%%%%%%%%%%%%%%%%%%%%%%%%%%%%
%  \vspace*{-0.5cm}
\begin{figure}[hbt]
\begin{center}
\includegraphics[width=8.5cm]{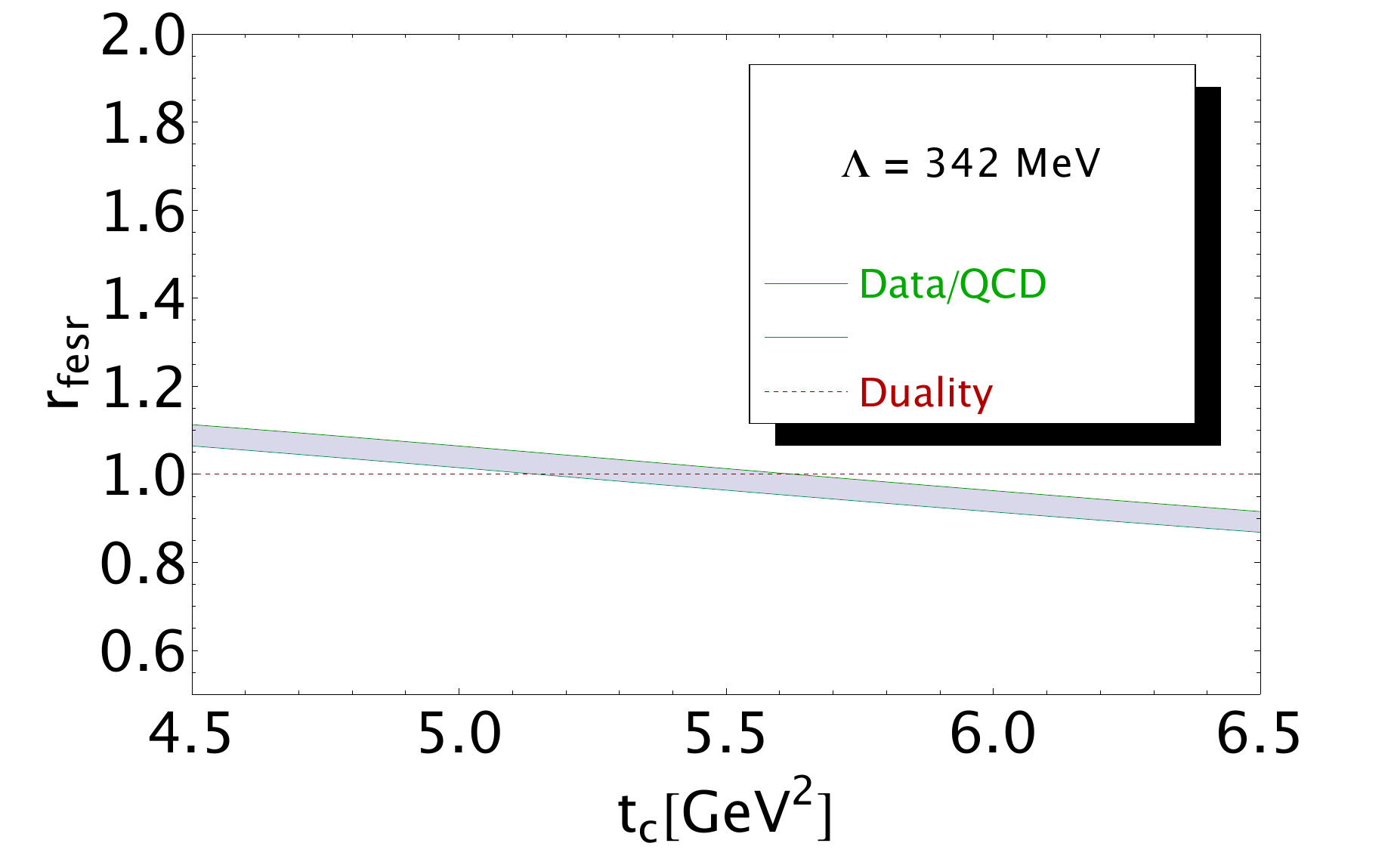}
\caption{\footnotesize Value of the ratio in Eq.\,\ref{eq:rfesr}.}\label{fig:s0}
\end{center}
\vspace*{-0.5cm}
\end{figure}  
%%%%%%%%%%%%%%%%%%%%%%%%%%%%%%%%%%%%%%%%%%%

\hspace*{0.5cm}\d One can fix $\hat s_0\equiv t_c$ from  the lowest
 FESR moment\,\cite{FESR}:
 \beq
F_0(t_c)\equiv  \int_0^{\hat t_c} d\,t\,R_{ee}^{I=1} =\frac{3}{2} \hat s_0\Big{[} 1+ a_s+3.891\,a_s^2+11.228\,a_s^3+\cdots \Big{]}, 
 \eeq
which, for a  $\rho$-meson dominance, gives\,\cite{FESR,PERROTET,SNe}:
 \beq
 \hat s_0=1.5~{\rm GeV}^2,
 \eeq

\hspace*{0.5cm}\d  Using the $e^+e^-\to$ Hadrons data, until $\sqrt{t_c}\simeq$ 1.875 GeV, the duality condition requires\,:
 \beq
  \hat s_0\equiv t_c=(5.1\sim 5.6)~{\rm GeV}^2,
 \eeq
 as shown in Fig.\,\ref{fig:s0} from SN23\,\cite{SNe} where we plot the ratio\,:
 \beq
r^{fesr}\equiv \frac{F^{data}_0}{F^{qcd}_0},
\label{eq:rfesr}
 \eeq
where  $r^{fesr}=1$ corresponds to the duality region. 
This value does not support the one 1.6 GeV$^2$ obtained in\,\cite{PERROTET} using the $e^+e^-\to$ Hadrons data (notice that in Ref.\,\cite{FESR} a value around 2.1\,GeV$^2$ has been obtained). However,  
one may (intuitively) expect a value of $t_c$ above the data region.  

\hspace*{0.5cm}\d  One can inspect that with a such value of $\hat s_0$, the additional contribution due to Duality Violation becomes unobservable when one uses a complete data thanks to the exponential weight for its parametrization. 
 
\hspace*{0.5cm}\d  A critical analysis of the effects of DV for  values of $\hat s_0$  around 1.5 GeV$^2$ on the determination of the QCD condensates and $\alpha_s$ using $\tau$-decay ALEPH data is explicitly discussed in Ref.\,\cite{PICH2}. 
 
\hspace*{0.5cm}\d  One should  also notice  that power corrections in the Euclidean space estimated previously in Table\,\ref{tab:cond} up to $D=20$  do not show any exponential behavior which disfavors by {\it a duality analytic continuation} the presence of this oscillating DV term in the Minkowski space.
 
  %%%%%%%%%%%%%%%%%%%%%%%%%%%%%%%%%%%%%%%%%%%%
  \section{Summary}
   %%%%%%%%%%%%%%%%%%%%%%%%%%%%%%%%%%%%%%%%%%%%
   In this paper, I have updated the results obtained in SN23\,\cite{SNe}.
   
  {\bf 1.} By using the new measurement of Ref.\,\cite{MG2} of the positive charge 
  muon anomaly, I found about $3\,\sigma$ deviation of the Standard Model predictions.
    
 {\bf 2.} I have combined the uses of ratio of LSR and some high $\tau$-decay like moments to improve the estimates of the QCD condensates of dimension  $D\leq 12$ dimensions and give new estimates of the ones $D=14$ to 20.  Up to this order, I have not observed  any exponential or/and factorial blow up of their contribution to the two-point function. I confirm the violation of four-quark factorization by a factor $\rho\simeq 6.89$  and the value of the gluon condensate $\la\alpha_s G^2\ra$ (though less accurate) found from heavy quark channels and some other sources.
  
  {\bf 3.} Using these new values of the condensates, I extract from the lowest
BNP moment the value of $\alpha _{s}(M_{\tau})$ using the lowest BNP $\tau$-like moment. The results from $e^+e^-\to$ Hadrons data are given in Eqs.\,\ref{eq:as-foci} and \ref{eq:as-e+e-}.  They are in excellent agreement with the new PDG average. The one from the V component of $\tau$-decay are in Eqs.\,\ref{eq:as-foci-tau} and \ref{eq:tau}, where we notice that more precise determinations of the $D=6,8$ condensates have improved the accuracy of $\alpha_s(M_\tau)$ from $\tau$-decay compared to the previous ones in the literature quoted in Table\,\ref{tab:tau}. The mean of the two determinations from $e^+e^-$ and $\tau$-decay data is in Eq.\,\ref{eq:as-final} which is in excellent agreement with the new PDG average. 
  
 {\bf 4.} I discuss, in Sections 10 and 11,  some (eventual) contributions  beyond the SVZ expansion ($1/Q^2$, instantons and Duality Violation). Such effects are expected to be relatively small (see e.g.\,\cite{SNpower}). The excellent
agreement between the present predictions within the standard SVZ expansion and the latest PDG average should provide a strong upper bound on the size of these extra contributions. 

  \vspace*{-0.25cm}
%\newpage
%%%%%%%%%%%%%%%%
\section*{Acknowledgements}
%%%%%%%%%%%%%%%%
It is a pleasure to thank Toni Pich and Valya Zakharov for some correspondences. 
 \vspace*{-0.25cm}
 %%%%%%%%%%%%%%%%%%%%%%%%%%%%%
 %%%%%%%%%%%%%%%%%%%%%%%%%%%%%

\end{document}